\documentclass[aps,prx,twocolumn,floatfix,superscriptaddress]{revtex4-2}

\usepackage{amsmath}  % needed for \tfrac, \bmatrix, etc.
\usepackage{amsfonts} % needed for bold Greek, Fraktur, and blackboard bold
\usepackage{graphicx} % needed for figures
\usepackage{epstopdf}
\usepackage{gensymb}
\usepackage{color}

\begin{document}

\title{Acoustic manipulation of multi-body structures and dynamics} 
% title brainstorming
% Acoustically interacting many-body systems

\author{Melody X. Lim}
\affiliation{James Franck Institute, The University of Chicago, Chicago, Illinois 60637, USA}
\affiliation{Department of Physics, The University of Chicago, Chicago, Illinois 60637, USA}
\author{Bryan VanSaders}
\affiliation{James Franck Institute, The University of Chicago, Chicago, Illinois 60637, USA}
%\affiliation{Department of Physics, The University of Chicago, Chicago, Illinois 60637, USA}
\author{Heinrich M. Jaeger}
\affiliation{James Franck Institute, The University of Chicago, Chicago, Illinois 60637, USA}
\affiliation{Department of Physics, The University of Chicago, Chicago, Illinois 60637, USA}
\begin{abstract}
Sound can exert forces on objects of any material and shape.
This has made the contactless manipulation of objects by intense ultrasound a fascinating area of research with wide-ranging applications.
While much is understood for acoustic forcing of individual objects, sound-mediated interactions among multiple objects at close range gives rise to a rich set of structures and dynamics that are less explored and have been emerging as a frontier for research. 
We introduce the basic mechanisms giving rise to sound-mediated interactions among rigid as well as deformable particles, focusing on the regime where the particles’ size and spacing are much smaller than the sound wavelength.  
The interplay of secondary acoustic scattering, Bjerknes forces, and micro-streaming is discussed and the role of particle shape is highlighted. 
Furthermore, we present recent advances in characterizing non-conservative and non-pairwise additive contributions to the particle interactions, along with instabilities and active fluctuations.
These excitations emerge at sufficiently strong sound energy density and can act as an effective temperature in otherwise athermal systems. 
\end{abstract}
\maketitle

Our everyday experience instructs us that matter shapes sound.
Changes in the shapes of our mouths and vocal chords give shape to words, which echo from hard walls and are muffled by soft surfaces, without moving or changing the internal structure of those obstacles.
Increase the sound intensity, however, and sound can in fact shape matter, as discovered by August Kundt in 1865~\cite{kundt1868iii}.
Kundt observed that when the air inside a sealed horizontal glass tube was set into resonance with an external sound wave, powder that was initially scattered at random inside the tube was collected into small piles.
The spacing of these piles was set by the wavelength of the sound, indicating a direct connection between the longitudinal propagation of sound waves in the tube and forces acting on the powder.
This observation set off a flurry of theoretical activity in search of the mechanism -- how do sound waves produce forces on objects? 

%In our everyday experience with sound, it is commonly heard but rarely {\it felt}.
%In typical situations where sound impinges on an object the transfer of momentum from wave to obstacle is too weak to cause motion or internal reconfiguration.
%Sufficiently intense sound, however, can indeed exert forces as it interacts with objects, and those forces can be used to manipulate or reconfigure them.
%This changes when the sound becomes sufficiently intense to exert a net force that then can be used to levitate objects and manipulate their configuration.

%At the same time, the idea that waves carry not only energy, but also momentum, was just entering the scientific consciousness, particularly beginning with electromagnetic waves.
In parallel, Lord Rayleigh pointed out in a 1902 paper~\cite{rayleigh1902xxxiv} that ``it seemed to me that it would be of interest to inquire whether other kinds of vibration exercise a pressure, and if possible to frame a general theory of the action", and proceeded to calculate the momentum carried by a vibrating gas~\cite{rayleigh1905xlii}.
This work laid the foundation for the first calculations of what is now referred to as the ``acoustic radiation force"~\cite{king1934acoustic,yosioka1955acoustic,westervelt1957acoustic}: the momentum transferred from an impinging acoustic wave to a rigid particle. 

\begin{figure*}
\centering
\includegraphics[width = 1.5\columnwidth]{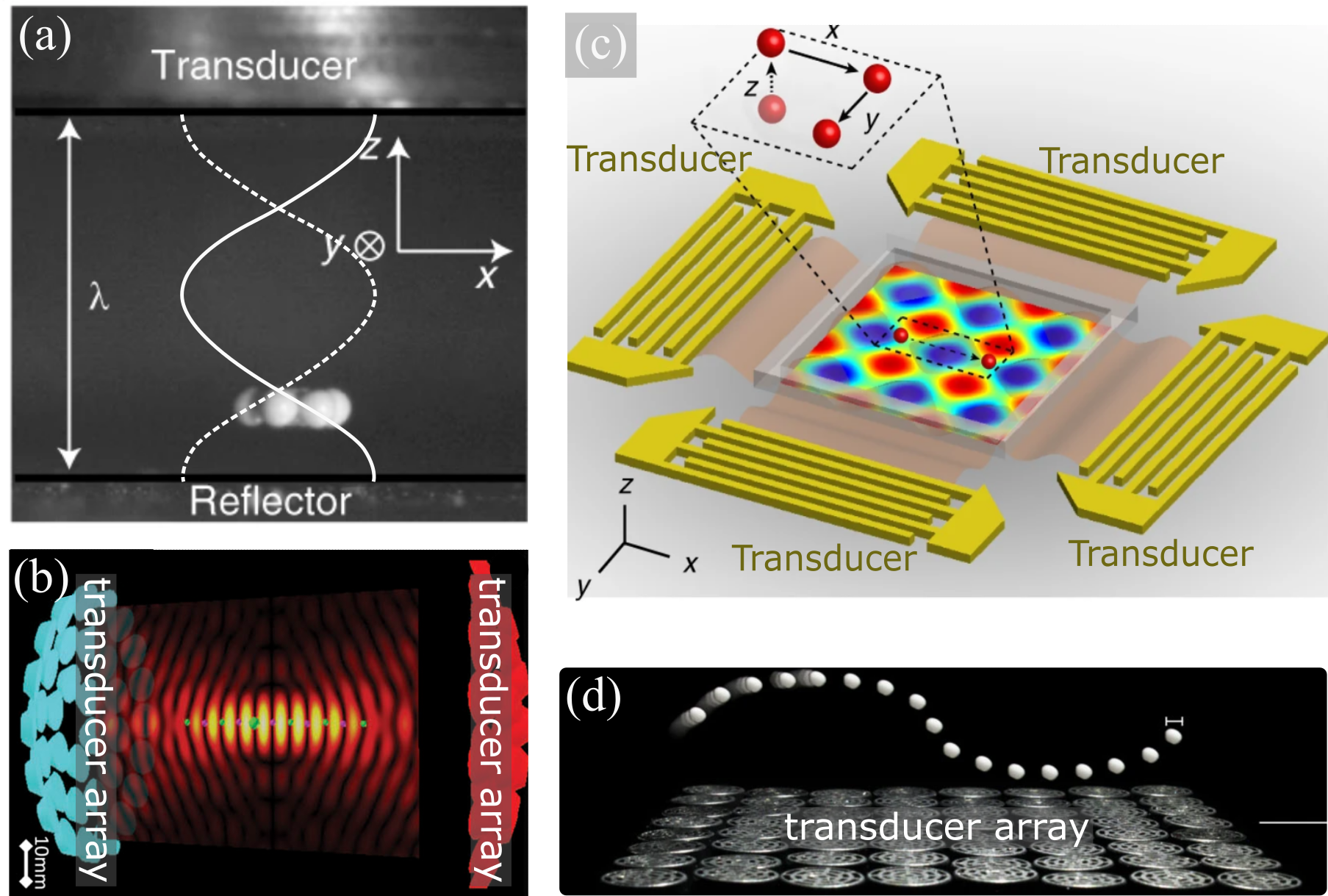}
\caption
{Experimental apparatus for manipulating particles with acoustic forces.
(a) \textcolor{black}{Side view of a} cluster of solid particles in a standing wave in air between a single transducer and a rigid reflector plate.
Particles levitate slightly below the node of the standing pressure wave (solid/dashed lines) due to gravity\textcolor{black}{, which points down along the vertical $z$-direction}.
Adapted from~\cite{lim2019cluster}.
(b) Simulated sound \textcolor{black}{pressure} field generated  \textcolor{black}{in air} by a pair of transducer arrays \textcolor{black}{shown in blue (left) and red (right)}.
Small objects can be levitated  \textcolor{black}{stably}  \textcolor{black}{in the nodal planes} at the locations  indicated, \textcolor{black}{irrespective of the orientation of the setup with respect to gravity}.
Adapted from~\cite{marzo2017tinylev}.
(c) \textcolor{black}{Perspective view of a} simulated  \textcolor{black}{two-dimensional} acoustic potential  \textcolor{black}{(well pattern)} generated by two  \textcolor{black}{orthogonal} transducer pairs in a microfluidic chamber.
The transducers excite a piezoelectric substrate, producing standing surface acoustic waves, which propagate into the  \textcolor{black}{liquid-filled central region. Controlling the transducers allows for particle transfer between wells (inset). Gravity points down along the $z$-direction}.
Adapted from~\cite{guo2016three}.
(d) Large single-sided transducer array, \textcolor{black}{seen from the side,} where the phase and amplitude of each element can be controlled to translate a polystyrene \textcolor{black}{ ball with diameter 2mm (scale bar indicated) from left to right}. \textcolor{black} {Gravity points downwards.}
Adapted from~\cite{marzo2019holographic}. }
\label{fig:primary}
\end{figure*}

Acoustic forces present a powerful platform for non-contact confinement and manipulation of objects of almost any material, and consequently have recently received growing attention.
In practice, the generation of acoustic forces for manipulating objects looks much like Kundt's original experiments: one or more sound sources (transducers) generate acoustic waves in a fluid, which produce acoustic forces on objects (Fig.~\ref{fig:primary}).
Generally, the required large sound intensity leads to the use of ultrasound frequencies, with characteristic wavelengths on the order of 0.1-10 millimeters.
In air, transducers are frequently used to excite modes of a resonant cavity, formed by the gap between a sound emitter and a reflecting surface (or another emitter), producing standing waves in which objects can levitate (Fig.~\ref{fig:primary}a,b).
A common alternative for objects immersed in liquids sees transducers excite standing surface waves in a piezoelectric substrate, which then radiate into a microfluidic chamber (Fig.~\ref{fig:primary}c).
\textcolor{black}{While we focus in this paper  on acoustic fields generated by standing sound waves, as in figure panels (a)-(c),} acoustically manipulating objects does not require a standing wave: a single-sided transducer array can also shape acoustic momentum so as to produce stable levitation 
 opposing gravity (Fig.~\ref{fig:primary}d). 

Each of the above setups relies on the ability of transferred acoustic momentum to contactlessly confine an object to a particular position.
%This is a single-particle effect: a single 
Any individual particle in an acoustic field experiences a force that draws it to a pressure node or an antinode, \textcolor{black}{depending on acoustic contrast between the particle and surrounding fluid}.
This single-particle effect, referred to as the primary acoustic force, is extremely powerful and general, and as such as been extensively leveraged for noncontact confinement and control in air (see recent reviews~\cite{andrade2018review,andrade2020acoustic,baudoin2020acoustic}).
In liquids and tissues, the ease and large depth with which high-frequency (MHz) ultrasound penetrates into soft materials has led to a wide range of biomedical and micro-fluidic device applications (for  reviews see~\cite{yeo2014surface,ozcelik2018acoustic,zhang2020acoustic}).
In each of these cases, work on radiation forces has focused on understanding and controlling the behavior of individual objects in an acoustic field. 

The same sound field that generates primary acoustic forces can also produce interactions among multiple objects, referred to as secondary acoustic forces.
These sound-mediated interactions that arise because of rescattering events from objects in an acoustic field, can be strong enough to affect the configuration and dynamics of groups of small particles.
As such, these interaction forces offer a wide range of additional opportunities to contactlessly assemble, aggregate, manipulate, and energize objects.
This is the case particularly in the regime of closely spaced, strongly acoustically interacting particles that forms the subject of our review.

As a general framework, we return to the question of how sound waves produce forces on objects.
The answer to this question is, ironically, the fact that matter shapes (the scattered) sound.
As a sound wave encounters an object, it emerges from the interaction with some altered momentum.
Momentum balance requires that a force is exerted on that object.
In the case of several objects, the rescattering of sound between particles produces additional acoustic forces.

\begin{figure}
\centering
\includegraphics[width = 0.85\columnwidth]{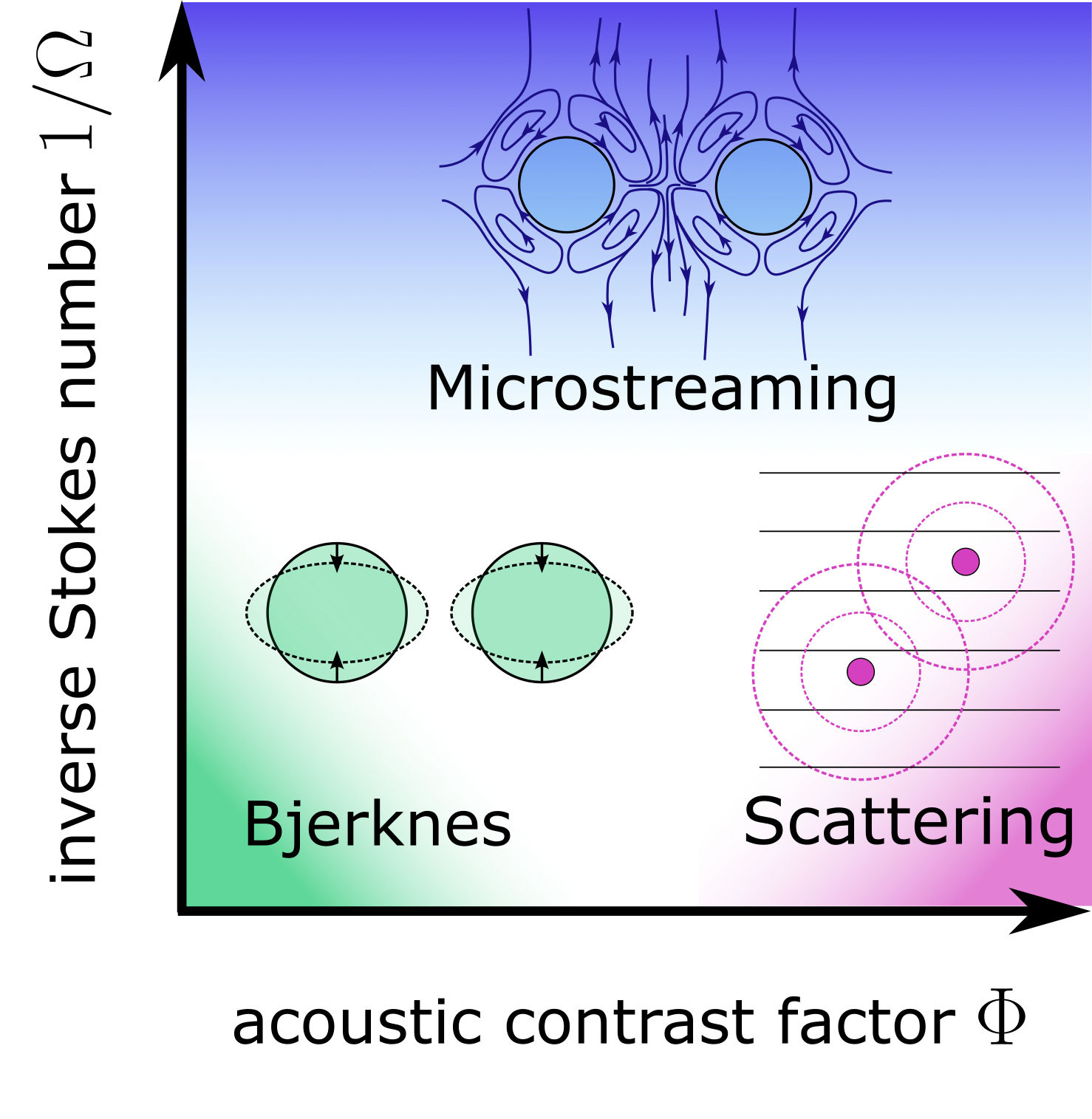}
\caption
{Regimes of multi-body acoustic interactions \textcolor{black}{within a levitation plane}.
%Particles with radius~$a$ and compressibility~$\beta$ are subject to a sound wave with frequency~$\omega$, which propagates in a fluid with kinematic viscosity~$\nu$ and compressibility~$\beta_0$.
\textcolor{black}{The acoustic contrast factor $\Phi$ determines whether levitated objects are moved to pressure nodes or anti-nodes by the primary acoustic force (acoustophoresis). It also controls the interactions among several objects within such nodes or anti-nodes, which are due to  sound scattered from the objects (secondary acoustic forces).
$\Phi$ has its maximum value 5/2 for incompressible objects much denser than the fluid medium, and it decreases as levitated objects become more compressible and/or less dense. 
In the limit where $\Phi \ll 0$ and levitated objects can sustain sound-induced shape oscillations, such as air microbubbles in water, Bjerknes forces become relevant for the interactions.
The relative importance of the medium's viscous dissipation is quantified by the inverse of the Stokes number~$\Omega$ %For two spheres micro-streaming starts to affect their interactions once $1/\Omega > 1/20$ \cite{fabre2017acoustic}
}
(schematic of flow lines and vortices surrounding two adjacent spheres adapted from~\cite{fabre2017acoustic}).
} \label{fig:regimes}
\end{figure}

In this review, we focus on three such mechanisms for producing particle-particle interaction forces \textcolor{black}{in an acoustic field}, which can be delineated according to the regime map presented in Fig.~\ref{fig:regimes}. These three mechanisms are the scattering of sound from objects which do not change shape under the action of sound (``scattering" regime), the oscillatory deformation of soft objects in response to an applied sound wave (``Bjerknes" regime), and the sound-induced flow of fluid around objects (``microstreaming" regime). These three regimes are delineated by the two quantities~$1/\Omega$ and~$\Phi$ on the axes of Fig.~\ref{fig:regimes}, which we now discuss in technical detail.

 \textcolor{black}{ When sound generates a force on a particle, the magnitude and direction of that force depends on the properties of the particle - specifically the differences in density $\rho$ and compressibility $\beta$ between the particle and the surrounding medium. These differences typically are expressed in terms of two coefficients,}
\begin{align}
f_0 &=1-\beta_p/\beta_m=1-\frac{c_m^2\rho_m}{c_p^2\rho_p} \nonumber\\ 
f_1 &=\frac{2(\rho_p/\rho_m-1)}{2\rho_p/\rho_m+1}.
\label{eq:f0f1}
\end{align}
\textcolor{black}{ Here the material densities of the fluid medium and the particle are $\rho_m$ and $\rho_p$, and the isentropic compressibilities of particle and medium are related to the associated speeds of sound ~$c_m$ and ~$c_p$ through $\beta_\mathrm{p}=(\rho_\mathrm{p}c^2_\mathrm{p})^{-1}$ and $\beta_\mathrm{m}=(\rho_\mathrm{m}c^2_\mathrm{m})^{-1}$.
A key parameter that determines not only the primary force on individual particles but also the interactions among particles via secondary scattering forces is the so-called acoustic contrast factor $\Phi$, given by}

\begin{equation}
    \Phi = f_0 + \frac{3}{2}f_1 = \frac{5\rho_p - 2\rho_m}{2\rho_p + \rho_m} - \frac{\beta_p}{\beta_m}.
\label{eq:Phi}
\end{equation}

This acoustic contrast factor has important implications not only in how individual particles couple to the sound field, but also for multi-body acoustic interactions.
%and we here use it to distinguish two regimes in multi-body acoustic interaction (plotted along the bottom axis of Fig.~\ref{fig:regimes}).
As concrete examples of different~$\Phi$, we can look at several common materials used for acoustic manipulation. For most objects levitated in air, including those made of rigid materials but also liquid droplets, the acoustic contrast factor approaches its maximum value, $\Phi = 5/2$, since their compressibility and density is sufficiently different than the levitation medium. 
For objects in a liquid medium $\Phi$ can be smaller, with values of 0.53 for polystyrene spheres or 0.025 for silicone oil droplets in water. 
$\Phi$ can become negative for emulsion droplets, such as soybean oil droplets in water (-0.20) or liquid perfluorohexane droplets in a lipid solution (-4.3) \cite{shakya2022acoustically}.

\textcolor{black}{As we discuss in more detail later in this article, 
individual objects levitated by a standing plane wave are moved by the primary acoustic force, often also called the acoustophoretic force, toward a pressure node when $\Phi > 0$, toward a pressure anti-node when $\Phi < 0$, and do not respond to primary acoustic forces when $\Phi = 0$.
The sign of $\Phi$ furthermore determines whether the dominant secondary interaction forces among particles in the (nodal or anti-nodal) levitation plane are controlled by compressibility or density differences between particle and medium.} 

\textcolor{black}{For particles where ~$\Phi$ is positive or has only a small negative value, such as in the above examples, the impinging sound does not cause rapid changes in particle shape, such that the dominant form of acoustic momentum transfer is scattering (lower right hand corner of Fig.~\ref{fig:regimes}).} 
However, when both the compressibility and density of the levitated object are much smaller than those of the medium, the forces become more complex.
  In this limit,  where $\Phi$ is very strongly negative, such as for gas bubbles in a liquid, the high gas compressibility enables sound-induced shape oscillations of the micro-bubbles. In turn, these generate bubble-bubble interactions labeled Bjerknes forces (lower left corner of Fig.~\ref{fig:regimes}). 

The  discussion so far assumes that sound is not attenuated in the surrounding fluid medium, i.e.~that the fluid is inviscid.
Viscous damping provides an additional mechanism for acoustic momentum to couple to bulk fluid flow, a phenomenon known as acoustic streaming.
As the primary sound field oscillates the fluid back and forth with angular frequency $\omega$ along the particle-fluid interfaces, it produces steady microstreaming flows within a boundary layer of characteristic thickness $\delta=\sqrt{2\nu/{\omega}}$, where $\nu$ is the kinematic viscosity of the fluid.
When the particle size $a$ \textcolor{black}{shrinks and} approaches this characteristic scale, micro-streaming starts to dominate the particle-particle interactions at close approach.
This crossover is governed by the
\textcolor{black}{Stokes number \cite{fabre2017acoustic}} 

\begin{align}
 \Omega = \omega a^2/\nu = 2(a/\delta)^2.   
\label{eq:Omega}
\end{align} 
In Fig.~\ref{fig:regimes} we plot $1/\Omega$ along the vertical axis. In the upper part of the diagram this introduces, for all acoustic contrast factors $\Phi$,  a regime where viscous effects need to be accounted for in describing interactions.  

The three regimes sketched in Fig.~\ref{fig:regimes} provide a rich platform for exploring multi-particle physics where particle interactions can be tuned \emph{in-situ}.
As with colloids and dusty plasmas, the regime of strong correlations (i.e.~average interaction energy much larger than kinetic energy) is easily reached under room temperature conditions and with particles large enough (a few microns to millimeters) to be tracked individually.
In contrast to colloids and dusty plasmas, particle charge is not required to stabilize the system or vary interparticle forces. As a result, in acoustic systems the steady-state interparticle spacing can be changed from direct contact to distances of several particle diameters.
Furthermore, particles levitated in air are underdamped, with inertia playing an important role in the evolution of strongly-interacting assemblies.

The remainder of this article is organized as follows.
A first section provides background about the three regimes in Fig.~\ref{fig:regimes} and discusses connections between acoustic and optical forces due to scattering.
Section II then outlines current frontiers in the area of multi-particle acoustic interactions.
This includes a discussion of how non-spherical particle shape modifies the secondary scattering forces, an effect that can be used to direct particle assembly.
It also discusses the coupling between moving objects and the sound mode in a resonating cavity, which can give rise to energizing instabilities that make it possible to drive levitated objects with an effective temperature. 
Finally, this section introduces under what conditions non-conservative and non-pairwise forces can be observed. 
With acoustic forces such conditions are accessed comparatively easily, which makes acoustic levitation a highly suitable platform for their investigation.
In Section III we survey applications based on the secondary radiation forces discussed in the preceding sections.
These range from micro-bubble aggregation for medical imaging to controlled assembly of objects inside microfluidic systems to large scale applications in liquid food or slurry processing.
We conclude with an outlook in Section IV, where we point to some of the outstanding challenges in modeling but also emphasize the unique opportunities offered by acoustic forces for exploring the physics of strongly correlated many-particle systems.

\section{Regimes of multi-body acoustic interactions} \label{major:one}

\subsection{\textcolor{black}{Scattering Forces on Objects without Sound-Induced Deformations}}
\label{sec:rigid}
We begin with what is perhaps the most straightforward way in which momentum from sound waves can be transferred to an object: scattering.
This type of acoustic force, referred to generally as the acoustic radiation force, is the dominant form of momentum transfer for rigid particles in a sound wave when the liquid viscosity can be neglected (lower right hand corner of Fig.~\ref{fig:regimes}).
Conceptually, the acoustic radiation force is the difference in momentum between the incoming and scattered acoustic waves, integrated over the particle surface.
As such, this regime has received a considerable amount of theoretical attention, which we briefly review here. 

In general, theoretical treatments of the acoustic radiation force are challenging because these forces are fundamentally second-order (i.e. nonlinear) effects.
To see this, we consider a sound wave with frequency~$\omega$. 
%(and corresponding wavelength $\lambda_0$ and wavenumber $k_0 = 2\pi/\lambda_0$).
Such a wave consists of  pressure~$p$, velocity~$\mathbf{v}$ (vectorial quantities are denoted with boldface type), and density~$\rho$ fields.
These fields have a spatial part (denoted here as a dependence on vector~$\mathbf{r}$), as well as a dependence on time~$t$.
Assuming that the sound wave is harmonic in time and perturbs a background \textcolor{black}{fluid medium} with rest pressure, density, \textcolor{black}{and sound speed}~$p_{m,0}$, ~$\rho_{m}$, and  $c_m$, respectively, we have

\begin{align}
    p(\mathbf{r},t) &= p_{m,0} + p(\mathbf{r}) e^{-i\omega t} \nonumber\\
    \mathbf{v}(\mathbf{r},t) &= \mathbf{v}(\mathbf{r}) e^{-i\omega t} \nonumber \\
    \rho(\mathbf{r},t) &= \rho_{m} + \rho(\mathbf{r}) e^{-i\omega t} \, .
    \label{eq:acfields}
\end{align}
Additionally, these \textcolor{black}{fields} are related through the velocity potential~$\phi(\mathbf{r},t) = \phi(\mathbf{r})e^{i\omega t}$, such that 

\begin{align}
    p(\mathbf{r}) &= i\omega \rho_{m} \phi(\mathbf{r}) \nonumber \\
    \mathbf{v}(\mathbf{r}) &= \nabla \phi(\mathbf{r}) \nonumber \\
    \rho(\mathbf{r}) &= i \frac{\omega \rho_m}{c_m^2}\phi(\mathbf{r})\, .
    \label{eq:def_vpot}
\end{align}
Considering these expressions, we arrive at an important consequence: the time-average of a purely harmonic pressure variation around $p_{m,0}$ is identically zero, and so to lowest order there is no net momentum transferred from the sound to a particle. Thus, the acoustic radiation forces must arise from additional nonlinear terms due to the presence of the particles in the acoustic wave. 

Computing these higher-order scattered acoustic waves, and the associated acoustic radiation forces, is not straightforward. Such a scattering problem depends, for instance, on the details of the particle geometry~\cite{fox1940sound,hasegawa1988acoustic,mitri2005acoustic}, whether or not the particles are compressible relative to the fluid~\cite{doinikov2001acoustic,mitri2005acoustic,mitri2006calculation,leao2016core}, and the arrangement of particles in the acoustic field~\cite{embleton1962mutual}. 

Progress can be made in certain limits. In 1934 L. V. King began by assuming incompressible particles and inviscid fluid~\cite{king1934acoustic}, which was later extended to include the effects of compressibility~\cite{yosioka1955acoustic,hasegawa1969acoustic,hasegawa1979acoustic}. Such methods generally write the acoustic radiation force as a sum of terms in a multipole expansion. Compact expressions can be derived by making the additional approximation that particles are compressible, but do not change shape in response to the applied acoustic field, and have a radius~$a$ much smaller than the wavelength of sound, a regime referred to as the Rayleigh limit. Since the particles are much smaller than the sound wavelength $\lambda$, anisotropy in the particle shape is much smaller than the diffractive limit on the features that can be resolved, and all particles can be treated as spherical. In this limit, the acoustic radiation force~$F_\mathrm{rad}$ on a point scatterer is conservative~\cite{Gorkov1961,Doinikov1994,Bruus2012,Settnes2012,silva2014acoustic}, and can be expressed as the gradient of an acoustic potential~$U_\mathrm{rad}$: 
%\textcolor{black}{Where is $\Phi$ here? You do not have it here: the trick is to look at when U is minimum (pressure node or anti-node)...and then you see that corresponds to the value of $\Phi$ (pull out a factor 1/2 and the prefactors for the two energy density terms are exactly what's in $\Phi$)}:

\begin{align}
F(\mathbf{r})_\mathrm{rad}&= -\nabla U_\mathrm{rad} \nonumber \\ 
U(\mathbf{r})_\mathrm{rad}&=\frac{4\pi}{3}a^3 \left[\frac{1}{2 }\textcolor{black}{\beta_m}f_0 \langle p(\mathbf{r},t)^2\rangle - \frac{3}{4}f_1\rho_m\langle |\mathbf{v}(\mathbf{r},t)|^2\rangle\right] \, ,
\label{eq:Urad}
\end{align}
where angled brackets denote averages over one acoustic cycle. This lowest-order expansion, introduced in 1962 by L. Gor`kov \cite{Gorkov1961}, separates the contribution of the pressure and velocity fields, and couples them to the acoustic potential via the scattering coefficients~$f_0$ and~$f_1$ \textcolor{black}{we introduced in Eq.~\ref{eq:f0f1}}. 
%which are the compressibility and density contrast, respectively:
%\begin{align*}
%f_0&=1-\beta_p/\beta_m=1-\frac{c_m^2\rho_m}{c_p^2\rho_p}\\
%f_1&=\frac{2(\rho_p/\rho_m-1)}{2\rho_p/\rho_m+1}.
%\end{align*}
%Here, sound propagates in the particle with speed~$c_p$, the particle material density is~$\rho_p$, and the \textcolor{black}{isentropic compressibilities of the particle and the fluid are related to the speed of sound through $\beta_\mathrm{p}=(\rho_\mathrm{p}c^2_\mathrm{p})^{-1}$ and $\beta_\mathrm{m}=(\rho_\mathrm{m}c^2_\mathrm{m})^{-1}$, respectively.~$f_0$ and~$f_1$ are often referred to as the monopole and dipole scattering coefficients respectively. }

Given that Eq.~\ref{eq:Urad} produces accurate predictions for the acoustic radiation force on a particle, the problem of finding the acoustic radiation force acting on that particle reduces to the problem of finding the fields~$p$ and~$v$, and substituting into Eq.~\ref{eq:Urad}. In turn, the forms of the pressure and velocity fields are a function of the boundary conditions under which the acoustic field propagates. In the case that the pressure and velocity fields correspond to those of an acoustic cavity, without the presence of any other particle, the acoustic force acting on the particle is referred to as the primary acoustic force. Adding another rigid particle as a boundary condition, computing the pressure and velocity fields due to the presence of that source particle, and then substituting into Eq.~\ref{eq:Urad}, produces the secondary acoustic force: the force on one particle due to the presence of another. 
This secondary acoustic force, which arises from rescattering events between particles in an acoustic field, can be computed analytically using perturbation expansions of~$p$ and~$\mathbf{v}$~\cite{silva2014acoustic,sepehrirahnama2022acoustofluidics,sepehrirhnama2020generalized}.

In the following, we focus on situations where sound pressure of amplitude $p_{m}$ excites standing waves along one direction, which we take as the $z$-direction.  \textcolor{black}{To gain some intuition as to the different possible behaviors, we consider first the primary acoustic force, which acts on a single particle placed in a standing wave. Such a particle will move towards a minimum in the acoustic potential  in Eq.~\ref{eq:Urad}. For a particle with $f_0$ and $f_1$ both positive, Eq.~\ref{eq:Urad} is minimized when~$\langle p^2 \rangle $ is zero, and this particle will move to an acoustic pressure node. On the other hand, if~$f_0$ and~$f_1$ are negative,~$U_\mathrm{rad}$ will be minimized instead when~$\langle v^2 \rangle$ is zero, and this particle will move to a pressure anti-node. The more precise distinction between particles that move to a node and anti-node is summarized by whether the previously introduced acoustic contrast factor,~$\Phi = f_0 + \frac{3}{2}f_1$, is positive or negative.}

Using Eq.~\ref{eq:Urad}, the form of the primary force on a levitated particle in a standing plane wave can then be described with the following expression:
\begin{align}
F(\mathbf{r})_\mathrm{rad}&= -\nabla U_\mathrm{rad} = -\frac{4\pi}{3}a^3 k E_0 \Phi \sin(2kz)\hat{\mathbf{z}},
\label{eq:F_prim}
\end{align}
where $k=2\pi/\lambda$ is the wavenumber, $E_0 = \frac{1}{2}\beta_m p_m^2$ the acoustic energy density in the cavity that forms the standing wave, and $z$ is distance from a pressure node.
%measures the excursion from the zero-force position.
%The offset $h$ is set to zero at a pressure node and to $kh = \pi/2$ at a pressure anti-node.
%\textcolor{black}{We could add here a sentence a la what Tom suggested about what sound energy density $E_0$ is required to levitate a small particle in air: "To give a concrete example, levitating 200 micron diameter polyethylene spheres in air at 40kHz, as in Fig. 1a, a sound energy density $E_0 \simeq xx$Pa is required". }

From Eq.~\ref{eq:F_prim} we see that the primary acoustic force is a restoring force that, for small excursions $z$ above and below the levitation plane, acts like a linear spring with stiffness proportional to the particle volume.
Particle motion is damped only by viscous drag from the surrounding medium.
In an underdamped system (e.g. when a particle levitates in air), this can lead to pronounced oscillations about the equilibrium position \cite{andrade2014experimental,lee2018collisional}.

\textcolor{black}{Once driven to the nodal or anti-nodal plane by the primary, acoustophoretic force, a levitated particle now experiences secondary acoustic forces due to the presence of other nearby particles.
For a pair of identical, completely rigid spheres in air, these interactions can already be quite complex (see Fig. 3).
The interaction due to scattering is attractive within the nodal plane ($z$ = 0), but repulsive when the second particle approaches the first from above or below (Fig. 3a), thereby biasing the formation of close-packed monolayer particle `rafts' within the nodal plane \cite{wang2015crystallization,lim2019cluster,lim2022mechanical}.}
%\textcolor{black}{However, micro-streaming can set up vortices just outside the particle surfaces, which produce repulsive interactions at close range (Fig. 3(b)).
%For sufficiently small Stokes number $\Omega$, i.e., in the upper regime of Fig. 2, this then leads to a competition between attractive forces due to scattering and repulsive forces due to micro-streaming \cite{wu2022hydrodynamic,fabre2017acoustic}.}

We next discuss these interaction forces in more detail, focusing first on secondary scattering when the acoustic contrast factor $\Phi$ is either positive or negative \textcolor{black}{in the absence of  sound-induced shape oscillations}.
%\textcolor{black}{In section \ref{sec:bjerknes} we examine the highly compressible limit, applicable to gas bubbles (so-called Bjerknes forces).
%Finally, streaming forces due to the viscosity of the surrounding fluid are discussed in \ref{sec:streaming}.}
%look at the limit that the particles are gas bubbles and thus extremely compressible (Bjerknes forces)} and then introduce micro-streaming.} 

\subsubsection{\textcolor{black}{Positive acoustic contrast}}
Positive acoustic contrast, in which case all particles will collect in a nodal plane, requires that $\Phi = f_0 + \frac{3}{2}f_1 > 0$. 
In practice, this is the situation not only for solid particles levitating in air, for which the set of scattering coefficients $(f_0, f_1)$ can be well approximated by (1,1), but also for
liquid droplets in air, so long as their shape remains approximately constant (large shape oscillations, breakup and coalescence introduce considerable complications \cite{hasegawa2020coalescence, abdelaziz2021ultrasonic}). 
$\Phi$ is also positive for solid spheres in water, such as polystyrene particles with $(f_0,f_1)=(0.47, 0.038)$, and for many types of live cells and liquid droplets in water, e.g. silicone oil droplets with $(f_0,f_1)=(-0.08, 0.07)$.

\begin{figure*}
\centering
\includegraphics[width = \textwidth]{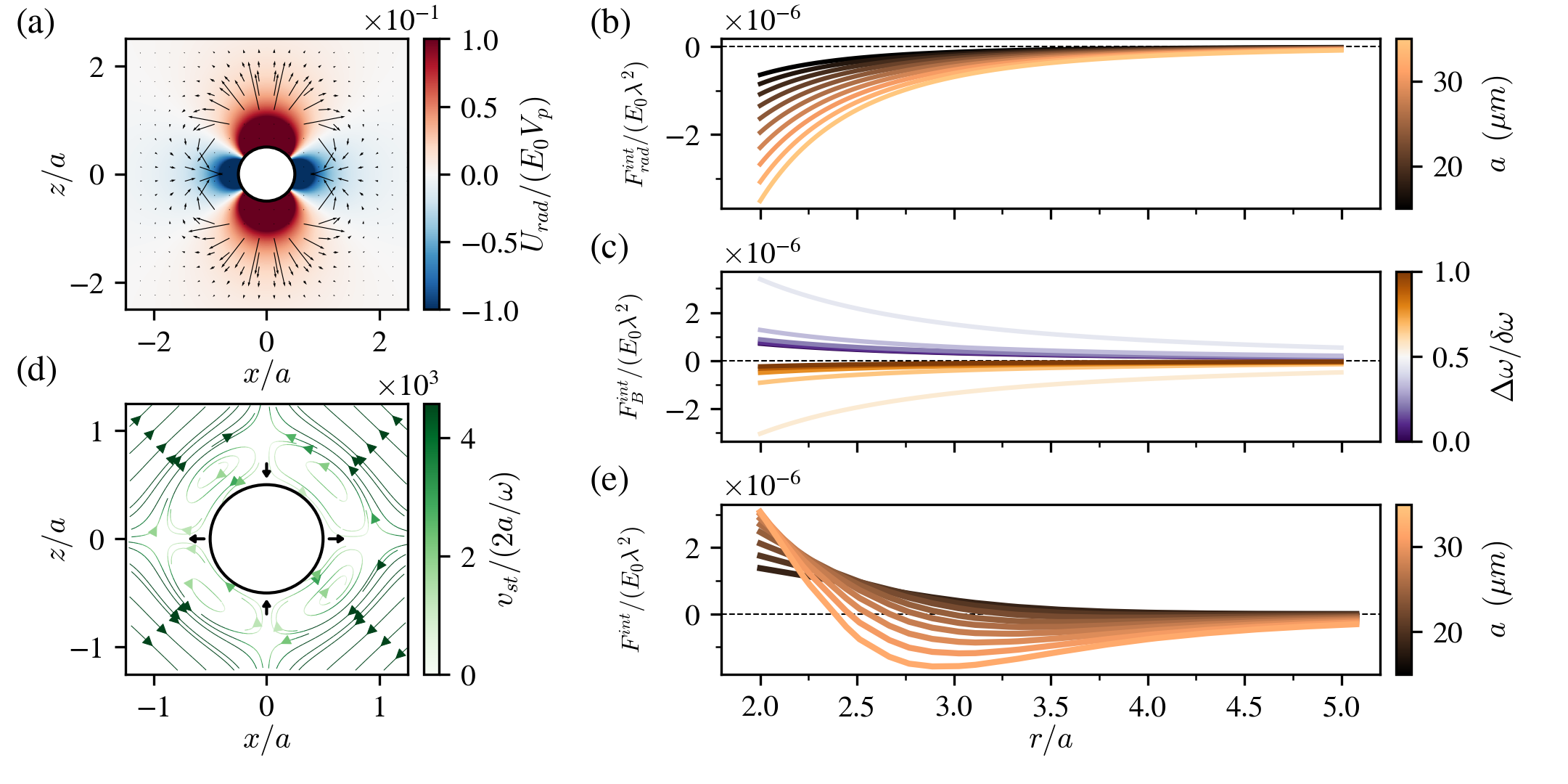}
\caption
{Acoustic pair forces.
(a) Acoustic potential (color) and secondary scattering forces (black arrows) experienced by a rigid sphere of radius $a$ approaching another, identical sphere centered in a pressure node at $z=0$. The image has azimuthal symmetry around an axis through $x=0$.  The acoustic potential $U_{rad}$ has been normalized by the sound energy density $E_0$ and the particle volume $V_p$.
%Pairwise scattering secondary forces between two identical, rigid spheres in the same nodal levitation plane.
%Color map shows the value of the acoustic potential, black arrows show the direction and relative magnitude of forces.
%Analytical form derived in \cite{silva2014acoustic}.
(b) \textcolor{black}{Radial interaction force due to} secondary scattering between two rigid, identical spheres within the same pressure node, \textcolor{black}{according to Eq.~\ref{eq:pairwise_Frad}} (in air, sound frequency 30 kHz).
Sphere radius is indicated by line color.
The plots give the radial force for $z=0$ in (a), normalized by the  energy density $E_0$ of the incident sound wave and its wavelength $\lambda$.
%As the diameter of particles is increased, attractive forces become more intense.
(c) Pairwise Bjerknes forces between two air bubbles (Eq.~\ref{eq:bjerk_secondary})
for a range of driving  frequencies $\omega$; the  bubbles have individual resonances centered on $\omega_m$ but $\delta \omega$ apart.
The driving frequency is varied as $\omega_m+\Delta\omega$.
For $\Delta \omega/\delta \omega <0.5$ interactions are repulsive.
Outside this window, interactions become attractive.
Sound frequency is 18.75 MHz, and both bubbles have resting radii on the order of $1\mu m$.
(d) Micro-streaming flows around a rigid sphere centered in a pressure node at $z=0$.
Streamlines are colored by relative velocity magnitude.
View is focused on the short-ranged `inner' set of vortices near the sphere surface.
Arrows indicate direction of flow of the inner vortices near the poles and equator of the sphere.
Analytical form derived in \cite{lane1955acoustical}.(e) Radial force between two identical, rigid spheres in a pressure node, similar to (b) but accounting for the viscosity of air  (sound frequency 30 kHz).
%Sphere  radius $a$ is indicated by line color.
As the sphere radius $a$ shrinks (line color), the Stokes number $\Omega$ decreases and interactions become progressively more repulsive as the result of competition between scattering and micro-streaming.
Adapted from \cite{wu2022hydrodynamic}.
}
\label{fig:forces}
\end{figure*}

\textcolor{black}{Using a perturbation expansion for the pressure and velocity fields in Eq. 5 the interaction force due to scattered sound between two particles can be calculated. 
For positive acoustic contrast this secondary acoustic force between two spheres of equal radius $a\ll\lambda$  levitating in the nodal plane with center-to-center distance $r\ll\lambda$ has, to lowest order, the form~\cite{silva2014acoustic,garcia2014experimental,hoque2020interparticle,lim2019cluster,hoque2021dynamical}}
\begin{align}
    F_{\mathrm{rad}}^{\mathrm{int}} (r) &= -\frac{3\pi E_0 a^6}{r^4}f_1^{(1)}f_1^{(2)} \ . 
    \label{eq:pairwise_Frad}
\end{align}
\textcolor{black}{We see that in the acoustic pressure node the compressibility becomes irrelevant and the secondary acoustic force therefore depends only on the density contrast via $f_1$.}
\textcolor{black}{The $r^{-4}$ dependence ensures that this in-plane force becomes significant only at close approach, typically a few particle diameters (Fig.~\ref{fig:forces}b). At the same time, the prefactors associated with this expression create several opportunities for the design of near-field acoustic interactions. First, $F_{\mathrm{rad}}^{\mathrm{int}}$ is negative and thus strictly attractive for particles of the same density (and thus same $f_1$), but can become repulsive when the two particles have opposite sign of ~$f_1$ (for example, if for two particles in a liquid one of them is slightly denser and the other slightly less dense than that liquid). In contrast to the primary force $F_{\mathrm{rad}}$ on each individual particle, which scales with $a^3$, i.e. with the particle volume, $F_{\mathrm{rad}}^{\mathrm{int}}$ between two spheres scales with the product of the two spheres' volumes, since the acoustic scattering event that creates the force involves interactions with both. Finally, increasing the acoustic energy density $E_0$ (experimentally, by increasing the amount of power injected to the acoustic cavity) linearly increases the magnitude of the secondary acoustic force. }

In the limit where~$r > \lambda$ (the far-field limit), the secondary acoustic force takes on the form
\begin{align}
    F_{\mathrm{rad}}^{\mathrm{int}}(r) = 2\pi E_0 k^2 a^6 \frac{\cos(kr)}{r^2}f_1^{(1)}f_1^{(2)}
        \label{eq:pairwise_Frad_far}
\end{align}
This long-range secondary interaction is oscillatory, indicating that there are acoustic potential minima, spaced at integer multiples of~$\lambda$ away from the central particle, where a second particle can be stably levitated~\cite{wang2017sound}. The depths of these potential minima again scale with the acoustic energy density $E_0$ and the product of two particle volumes. Unlike the close-range acoustic force Eq.~\ref{eq:pairwise_Frad}, the magnitude of long-range secondary interactions depends on the wavenumber~$k$, raising possibilities for separate tuning of the near- and far-field radiation force landscapes for acoustically levitated particles. 

As long as the point particle approximation remains valid,  the net acoustic force on a single particle due to several others can be expressed as the pairwise sum of the secondary acoustic forces due to all other particles~\cite{silva2014acoustic,zhang2016acoustically}: 

\begin{align}
    F_\mathrm{tot,i}^{\mathrm{int}} = \sum_{i\neq j} F_\mathrm{i,j}^{\mathrm{int}}
\end{align}

 For this regime mean-field theories, such as the one developed by Silva and Bruus \cite{silva2014acoustic} \textcolor{black}{or Sepehrirahnama and coworkers \cite{sepehrirhnama2020generalized,sepehrirahnama2022acoustofluidics}}, provide accurate predictions.
 Other regimes, however, are highly relevant and in need of exploration for cases where particles can no longer be treated as a point particle. This can happen in several senses. First, for dense particle configurations, where the spacing $r$ gets close to the particle size, the point-particle assumption no longer remains accurate, and furthermore details of the particle shape start to play a significant role, as will be discussed below.   
 Second, a particle may no longer be small compared to~$\lambda$. Significant departures from the Gor'kov theory begin to take place for spherical particles with diameter larger than~$0.3\lambda$. For such larger particles,  generally referred to as Mie particles, the scattered field can no longer be treated as small compared to the background acoustic field that is incident on the particles, rendering the perturbation theory approach that led to Eq.~\ref{eq:Urad} invalid. 
 In the most extreme case, a particle may be large enough to  preclude the formation of an acoustic standing wave. Additionally, since Mie particles can no longer be treated as pointlike relative to the acoustic field, detailed calculations of the acoustic forces acting on a levitated particle will depend  on the shape and acoustic excitations inside the particle. 

 The total acoustic radiation force acting on a levitated particle is the integral of the momentum flux over its surface, and so the problem of calculating the acoustic radiation force reduces to computing the scattering coefficients for an arbitrarily located object, which describe how the object couples to the basis wave fronts of the incident wave. The total force is then computed as the sum of a series of terms, which cannot necessarily be truncated because higher-order terms are not small compared to the incident fields. In the case of objects with a high degree of symmetry, such as spheres, analytical solutions to this scattering problem can be developed by analogy to the generalized Lorenz-Mie theory in optics~\cite{maheu1988concise,baresch2013three,lopes2015acoustic,sepehrirahnama2015numerical,marston2017finite,ospina2022particle}. Recent work has engaged in a detailed comparison of the primary levitation forces on a levitated object as a function of its size~\cite{ospina2022particle}. This work has shown that, for particular sizes, Mie particles can be stably levitated by plane-waves in pressure anti-nodes, but off-axis  (in contrast to Rayleigh particles, which levitate on-axis and in pressure nodes). More generally,  the stable levitation of Mie particles has focused, especially on the experimental side, on shaping the incident wavefront via computational methods~\cite{marzo2018acoustic,zehnter2021acoustic,tang2021acoustic,li2022holographic,hirayama2022high}.
 
 In addition to altering the stable levitation points (i.e., the effective primary acoustic forces) on a particle, increasing the particle size also changes the relative importance of the secondary and primary acoustic forces. Recent computational and experimental studies have shown that the secondary scattering force dominates for particles with diameter larger than roughly half a wavelength~\cite{silva2019particle}. For Mie particles smaller than this size limit, the secondary force varies in magnitude with the particle size compared to the primary force. As a result, in specific particle size ranges (particle diameter in the range of 0.28-0.31$\lambda$, as reported in~\cite{collins2015two}) the contribution of the secondary force towards particle clustering can be neglected, whereas the primary force remains strong, leading to the separation of Mie objects into individual acoustic wells~\cite{collins2015two,brugger2018orchestrating}.  Alternatively, the strong scattering from Mie particles can be used to create a series of traveling waves, as an alternative to particle aggregation~\cite{tang2021mie}. Such forces between bound clusters of Mie particles can lead to non-conservative forces, including structures that have driven degrees of freedom~\cite{clair2023dynamics,carvalho2022asymptotic}.

\subsubsection{\textcolor{black}{Negative acoustic contrast}}
For acoustic waves in liquid, it is possible for the acoustic contrast factor $\Phi$ in Eq.~\ref{eq:Phi} to become negative.
%We turn now to this case, represented in Fig.~\ref{fig:regimes} as reducing the compressibility ratio~$1-\beta/\beta_0$ along the bottom axis. 
%Acoustic radiation forces are the result of waves scattered from the boundary of objects immersed in a medium.
%Such reflections are of course only possible when the object is mechanically distinct from the surrounding medium, i.e.~when one or both the monopole and dipole coupling coefficients $f_1$ and $f_2$ are not zero.
%Objects with the same density and compressibility ($\kappa = 1/c^2\rho$) as their surrounding medium are invisible to passing pressure waves.
%For levitation in air, the density and compressibility differences between the surrounding fluid (air) and any solid are so large that both coupling coefficents are essentially equal to unity.
%This is not generally the case for bubbles immersed in fluids, where the compressibility ratio $\beta/\beta_0$ is larger than unity, causing $f_1<0$.
%To lowest order, the effect of compressibility can be included in Eq.~\ref{eq:Urad} by taking~$f_1<0$. 
As a result, the primary radiation forces which act on such objects are opposite in direction to those commonly observed to act on solid particles 
in air or even many types of cells and liquid droplets in water. Instead objects with $\Phi < 0$ are driven to pressure anti-nodes.
As mentioned earlier, examples are droplets of soybean oil in water and perfluorohexane droplets in lipid solution, for which the scattering coefficients $(f_0, f_1)$ take on the values (-0.11, -0.06) \cite{silva2014acoustic} and (-4.74, 0.32) \cite{shakya2022acoustically}, respectively.
\textcolor{black}{As before, we consider the case where particle shape change is negligible. This assumption is significantly violated for bubbles in a liquid medium, which have $f_0 \ll 0$ and therefore $\Phi \ll 0$ and can undergo large volume oscillations.
For this reason, special consideration must be given to the case of bubbles (see Section \ref{sec:bjerknes}).}
%The most common example are small gas bubbles in a liquid, for which the extremely negative values of the compressibility contrast $f_0$ dominate the overall acoustic contrast $\Phi$}. 

\textcolor{black}{In contrast to the physics at pressure nodes, where particle interactions involve only the density scattering coefficient $f_1$, when $\Phi < 0$ it is the compressibility ratio in $f_0$ that drives secondary interactions at pressure anti-nodes.}
As a result, the same perturbation analysis that led to Eqs.~\ref{eq:pairwise_Frad} and~\ref{eq:pairwise_Frad_far} now gives \cite{silva2014acoustic} 
\begin{align}
    F_{\mathrm{rad}}^{\mathrm{int}} (r) &= -\frac{4\pi}{9} \frac{E_0 k^2 a^6}{r^2}f_0^{(1)}f_0^{(2)} 
    \label{eq:pairwise_Frad_antinode}
\end{align}
for the in-plane near-field limit ($r \ll \lambda$) and
\begin{align}
    F_{\mathrm{rad}}^{\mathrm{int}} (r) = -\frac{4\pi}{9} E_0 k^3 a^6 \frac{\sin(kr)}{r}f_0^{(1)}f_0^{(2)}    
    \label{eq:pairwise_Frad_far_antinode}
\end{align}
for the in-plane far-field limit ($r > \lambda$). These interactions have smaller exponents in their power law decays and thus are longer ranged than their counterparts for the nodal plane. The near-field limit is still attractive for objects with the same scattering properties, and in the far field limit $F_{\mathrm{rad}}^{\mathrm{int}} (r)$ still oscillates and changes sign every half wavelength $\lambda/2$.  
\subsection{Bjerknes Forces}
\label{sec:bjerknes}
%Our discussion of acoustic radiation forces has so far focused on the regime where particles are rigid: that is, when waves scatter from the boundary of objects immersed in a medium without substantially changing the object shape.
%This important simplification allowed us to couple an object to the acoustic field by only its density and compressibility contrast factors (Eq.~\ref{eq:Urad}).
%For levitation in air, the density and compressibility differences between the surrounding fluid (air) and any solid are so large that both coupling coefficients are essentially equal to unity.
%As such, the secondary interactions of liquid droplets in air tend to have the same basic interactions as solid objects, with the obvious complication that object geometry is variable \cite{hasegawa2020coalescence, abdelaziz2021ultrasonic}. 

%\textcolor{blue}{HJ: I have been trying to connect the section on Bjerknes forces a bit more with what we write in the section on radiation forces above. Turns out that the primary Bjerknes force is of very similar form as the primary radiation force, and that the secondary Bjerknes force is the same as the secondary radiation force if the sound frequency is much smaller than the resonance frequency of the bubbles. Below is a modified front end for this section that takes this into account. Note that some of the equation numbering is off because I just added this new text while keeping the old one in place.
%=================}
When the compressibility of the objects subjected to the sound pressure becomes  large compared to the surrounding medium, additional physics can enter. This is specifically the case for gas bubbles: their size is a function of pressure within the medium, and so a passing pressure wave induces an oscillation in volume.
This volume oscillation then can itself radiate pressure waves.
The acoustic forces acting on small bubbles in a liquid are generally termed Bjerknes forces, after C.A.~Bjerknes and his son V.F.K.~Bjerknes \cite{bjerknes1906fields}.
The study of bubbles in applied sound fields has received considerable attention due to relevance in  scenarios including sonochemistry \cite{stricker2013interacting, shchukin2006sonochemical, fernandez2010efficient}, medical ultrasonic imaging \cite{dayton1997preliminary, lazarus2017clustering, harvey2002advances, blomley2001microbubble, garbin2011unbinding, kokhuis2013secondary, navarro2022monodispersity}, micromanipulation \cite{xu2013microbubble, goyal2022amplification}, directed transport within the bloodstream \cite{fonseca2022ultrasound} and cavitation \cite{mettin1997bjerknes}.
Various authors have contributed to refining the analytical description \cite{pelekasis1993bjerknes_a, pelekasis1993bjerknes_b, doinikov1995mutual, doinikov2003acoustic} or experimental record \cite{crum1975bjerknes, jiao2015experimental, yoshida2011experimental, rabaud2011acoustically, garbin2009history, zeravcic2011collective}.

When a bubble is placed within an ideal fluid with a harmonically oscillating pressure field $p=p_m \sin (\omega t)$, small deformations $\epsilon$ away from the equilibrium radius $a_0$ obey
\begin{equation}
\frac{\partial^2 \epsilon}{\partial t^2} + \omega_c^2 \epsilon = \frac{p_a}{\rho_m a_0}\sin(\omega t),
\end{equation}
where $\rho_m$ is the liquid density and $\omega_c$ is the resonant frequency of the bubble, given by
\begin{equation}
\omega_c^2 = \frac{1}{a_{0,c}^2 \rho_m}\left[3\kappa\left(p_{m,0} + \frac{2\sigma}{a_0}\right) - \frac{2\sigma}{a_0}\right] \, .
\label{eq:bubble-resonance}
\end{equation}

Here $p_{m,0}$ is the (constant background) hydrostatic pressure and $\sigma$ is the liquid surface tension.
The constant $\kappa$ is known as the gas polytropic index, which quantifies the deviation of the gas equation of state from ideal behavior ($p\propto \rho^{1+1/\kappa}$).
Equation 14 connects a bubble's resonant frequency to its size.
As for forced harmonic oscillators generally, bubbles with resonant frequencies $\omega_c$ significantly greater than the frequency $\omega$ of the applied pressure field (or equivalently, bubble sizes $a_0$ smaller than the resonant size $a_{0,c}$) oscillate in phase with the applied field.
Conversely, bubbles with significantly smaller resonant frequencies oscillate out of phase.

As a consequence, individual bubbles will migrate to a pressure anti-node for sound frequencies $\omega < \omega_c$ and to a pressure node for $\omega > \omega_c$. 
This is the primary Bjerknes force ~\cite{Leighton1990primary,doinikov2003acoustic}. 
The direction of the primary Bjerknes force on a bubble at a particular location $z$ along a standing acoustic wave can therefore be reversed as a function of the applied frequency, or alternatively by changing the bubble size.

For small deviations from a node or anti-node the primary Bjerknes force is a restoring force and can be expressed as
\begin{equation}
    F_{B} = - \frac{4 \pi a_0^3}{3} k E_0 \frac{\beta_p}{\beta_m} \frac{\omega_c^2}{\omega^2-\omega_c^2} \sin(2kz). 
    \label{eq:bjerk_primary}
\end{equation}
As before, $z$ is the distance measured from a pressure node.
%, and the offset is $kh =0$ for a pressure node and  $kh =\pi/2$ for an anti-node.
 The above expression for the primary Bjerknes force neglects dissipation due to factors such as fluid viscosity and heat conduction, and for a more complete description we refer the reader to the extensive analysis by A.~Doinikov in Ref.~\cite{doinikov2003acoustic}.
 \textcolor{black}{In the limit that~$\omega \ll \omega_c$, corresponding to either low sound frequency or very small bubbles, bubble volume oscillations are negligible, and Eq.~\ref{eq:bjerk_primary} reduces to  Eq.~\ref{eq:F_prim} if we use ~$\Phi \approx -\beta_p/\beta_m$, as appropriate for highly compressible objects at anti-nodes. }

Equation \ref{eq:bjerk_primary} expresses the primary force on bubbles due to \textcolor{black}{a pressure gradient arising from} the externally applied sound field.
\textcolor{black}{However, forces} arise in the presence of any pressure gradient, and so a `secondary' Bjerknes force also exists, which as in the radiation case discussed above is the reaction of a bubble to the scattered pressure field of another nearby bubble.

%\textcolor{black}{Consider two similar bubbles. When the sound frequency $\omega$ is much smaller than either bubble’s resonance frequency $\omega_c$ they will both move to the pressure anti-node and oscillate in phase with the sound pressure. 
%As shown by Zheng and Apfel \cite{ZhengApfel1995} and also derived by Silva and Bruus \cite{silva2014acoustic}, in this case the near-field ($2a_0 < r \ll \lambda)$ interaction between two identical bubbles is simply the secondary acoustic force due to scattering between objects with $\Phi \ll 0$, given by Eq. \ref{eq:pairwise_Frad_antinode}, where we can  approximate the product of contrast factors $f_0^{(1)}f_0^{(2)} \approx (\beta_p/\beta_m)^2$. } 

%\textcolor{black}{The situation becomes more complex for arbitrary bubble sizes and thus bubble resonance frequencies.} 
For the case of two bubbles in an inviscid fluid under a long-wavelength (i.e.~much larger than the bubble diameter) standing pressure wave, the following form of the secondary force was derived by Bjerknes and Bjerknes, which, following \cite{doinikov2003acoustic} but using our notation, can be written as

\begin{equation}
   F_{B}^\mathrm{int}(r) = -\frac{8\pi k^2 E_0}{9} \frac{a_{0,1}^3 a_{0,2}^3}{r^2} \frac{\beta_{p,1}}{\beta_m}\frac{\beta_{p,2}}{\beta_m}\frac{\omega_{c,1}^2 \omega_{c,2}^2}{(\omega_{c,1}^2-\omega^2)(\omega_{c,2}^2-\omega^2)}. 
   \label{eq:bjerk_secondary}
\end{equation}

Here $a_{0,i}$ are the equilibrium radii of the two bubbles, each with associated resonance frequency $\omega_{c,i}$ according to Eq.~\ref{eq:bubble-resonance}, and $r \ll \lambda$ is the distance between bubbles \cite{bjerknes1906fields, doinikov2003acoustic} \textcolor{black}{(we are again ignoring effects due to various dissipation mechanisms; see \cite{doinikov2003acoustic})}.
Equation \ref{eq:bjerk_secondary} reveals that bubbles repel ($F_B^{int}>0$) whenever $\omega$ lays between $\omega_{c,(1)}$ and $\omega_{c,(2)}$ \textcolor{black}{(in which case they also move to different nodes)}, and they attract otherwise, as confirmed by experimental observations \cite{crum1975bjerknes}.
\textcolor{black}{In Fig.~\ref{fig:forces}c we show numerical results for the interaction of two micro-bubbles according to Eq. \ref{eq:bjerk_secondary}.}
\textcolor{black}{In the limit where $\omega$ is much smaller than either of the bubbles' resonance frequencies, such that bubble shape oscillations become negligible,  Eq.~\ref{eq:bjerk_secondary} reduces to Eq.~\ref{eq:pairwise_Frad_antinode} for the secondary acoustic near-field interaction if we also recall that  $f_0^{(i)} \approx -\beta_{p,i}/\beta_m$ for bubbles \cite{ZhengApfel1995,silva2014acoustic}.}

However, when bubbles approach at very close range (i.e.~on the order of a bubble diameter or less), Eq.~\ref{eq:bjerk_secondary} is insufficient to explain bubble interactions.
The most striking effect in need of explanation was the reversal in secondary Bjerknes force at close range, such that bubbles attractive at longer ranges will approach but not coalesce, instead halting at a stable distance \cite{barbat1999dynamics, yoshida2011experimental,regnault2023dynamics, fernandez2010efficient}, forming clusters that have been called `bubble grapes.'
Numerous theoretical studies have approached this problem, extending the analysis to include the effects of sound being scattered between bubbles multiple times \cite{doinikov1995mutual}, couplings between the bubble oscillations \cite{barbat1999dynamics, ida2002characteristic,ida2003alternative,lanoy2015manipulating}, or effects due to the anharmonicity of bubble oscillations outside the harmonic regime \cite{mettin1997bjerknes}. 
Recent studies in shear-thinning media have even demonstrated complex surface modes on interacting bubbles, responsible for self-propulsion of bubble trains \cite{janiak2023acoustic}.

\subsection{Micro-Streaming}
\label{sec:streaming}
An additional and important close-range correction to the forces between bubbles are the streaming interactions \cite{doinikov2003acoustic}.
Generally, these are steady flows that arise in liquids under oscillating pressure fields, and are nonlinear in nature.
Due to the particularly high sound intensities used in sonochemistry or cavitation studies, streaming forces are important to the dynamics of groups of bubbles \cite{pelekasis2004secondary, doinikov1999bjerknes, marmottant2006high, zeravcic2011collective}.
However, streaming effects are possible in any acoustic context involving a non-inviscid fluid, and so are sufficiently general that we devote the following section to their consideration.

The viscosity of the host fluid provides another means for the generation of acoustic forces (upper regime in Fig.~\ref{fig:regimes}), specifically through the generation of steady micro-streaming flows. Although ordinarily this regime is associated with the increase of fluid viscosity (e.g., the propagation of sound in water rather than in air), viscosity can also play an appreciable effect for sufficiently small particles in air. As an example, we return to Kundt's observation of clustering powder in a resonant tube. Although the formation of these clusters was at first thought to be due to the vibration of the tube, or perhaps acoustic radiation forces, later experiments demonstrated that the air inside the tube in fact developed circulatory currents, which displaced the powder. 

Just thirty years earlier, Faraday~\cite{faraday1837peculiar} had observed a similar invisible flow during experiments on Chladni plates. When heavy grains (such as sand) are displaced on a Chladni plate, they move to the nodal lines (i.e. where the amplitude of harmonic displacement of the plates was zero). However, repeating the experiment with light powders produced clusters at the antinodes of the vibrating plate, where the displacement was greatest. Careful observation revealed that the powders collected in the air in ``parcels, which are in extraordinary condition: for the powder of each parcel continues to rise up at the centre and flow down on every side to the bottom, where it enters the mass to ascend at the centre again"~\cite{faraday1837peculiar}. The powder in fact collected at the antinodes because the currents of air rise from the areas of maximum vibration. 

These observations, as pointed out by Lord Rayleigh~\cite{strutt1884circulation}, indicate that harmonic motion (of either a plate or a tube) produces steady-state circulatory flow away from the vibrating solid surface. Such flows arise due to the effect of ``friction, by which the motion of fluid in the neighborhood of solid bodies may be greatly modified"~\cite{strutt1884circulation}, and are generally referred to as Rayleigh streaming. 
We  note that the propagation of intense sound waves in free space can produce steady-state fluid flow in the absence of solid boundaries, due to the gradient in radiation pressure along the beam propagation direction. Such streaming is referred to as Eckart streaming~\cite{eckart1948vortices, boluriaan2003acoustic, wiklund2012acoustofluidics}, and takes place over lengthscales much larger than the acoustic wavelength. Our subsequent focus is on Rayleigh streaming, or micro-streaming, in close proximity to particle surfaces. 

In order to gain some intuition as to the physical origin of steady-state micro-streaming flows, we turn to the Navier-Stokes equations for a \textcolor{black}{compressible fluid} in the absence of additional body forces~\cite{lighthill1978acoustic,bruus2012acoustofluidics2, sadhal2012acoustofluidics}

\begin{align}
    \nabla p &= -\rho\frac{\partial \mathbf{v}}{\partial t} - \rho(\mathbf{v}\cdot \nabla ) \mathbf{v} +\eta \nabla^2 \mathbf{v} + \zeta \eta \nabla(\nabla \cdot \mathbf{v}) \, .
\label{eq:navierstokes}
\end{align}
\textcolor{black}{Here the viscosity ratio $\zeta$ accounts for the internal friction of the fluid medium under compression, which differs from the internal friction in response to shear that is represented by $\eta$.  
Substituting the pressure, velocity and density fields from Eq.~\ref{eq:acfields} into Eq.~\ref{eq:navierstokes} and treating the oscillating terms (which we here denote by the subscript 1) as small perturbations gives, to first order in these perturbations, 
$\nabla p_1 = -\rho_m\frac{\partial \mathbf{v}_1}{\partial t} +\eta \nabla^2 \mathbf{v}_1 + \zeta \eta \nabla(\nabla \cdot \mathbf{v}_1)$.
Since the terms in this first order expression oscillate harmonically in time, there is no flow averaged over an acoustic cycle. The origin of acoustic streaming must therefore be in higher order terms. In order to find these terms, we operate under the assumption that there are second order perturbations to Eq.~\ref{eq:acfields}, labeled with subscript 2, which do not depend on time:}
\begin{align}
    p(\mathbf{r},t) &= p_{m,0} + p_1(\mathbf{r},t) + p_2(\mathbf{r}) \nonumber \\
      \mathbf{v}(\mathbf{r},t) &= \mathbf{v}_1(\mathbf{r},t)+ \mathbf{v}_2(\mathbf{r}) \nonumber \\
     \rho(\mathbf{r},t) &= \rho_m +\rho_1(\mathbf{r},t)+ \rho_2(\mathbf{r}) \, .
\end{align}

\textcolor{black}{Inserting these fields into Eq.~\ref{eq:navierstokes} and averaging all terms over one oscillation cycle yields, to second order and after rearranging \cite{bruus2012acoustofluidics2}}, 

\begin{align}
    \begin{split}
        -\nabla p_2 + \eta \nabla^2 \mathbf{v}_2 + \zeta \eta \nabla(\nabla \cdot \mathbf{v}_2) = \\
    \Bigl\langle \rho_1 \frac{\partial \mathbf{v}_1}{\partial t} \Bigr\rangle 
    + \Bigl\langle \rho_{m}(\mathbf{v}_1 \cdot \nabla )\mathbf{v}_1 \Bigr\rangle \, .
    \end{split}
\label{eq:streaming}
\end{align}

\textcolor{black}{While~$p_1$, ~$\mathbf{v}_1$ and~$\rho_1$ individually vary harmonically, and so average out to zero over an acoustic cycle, the product of two harmonic quantities does not generally average to zero over time. Thus, the two terms on the right hand side of Eq.~\ref{eq:streaming}, which are products of oscillating entities, produce a non-zero streaming velocity field $\mathbf{v_2}$ on the left hand side. 
Taking into account the appropriate boundary conditions, this gives rise to the velocity field surrounding the sphere shown in Fig. \ref{fig:forces}d. The same oscillating entities also generate a steady pressure, $p_2$, whose gradient provides the primary acoustic force that drives acoustophoretic motion.  
We further note that coupling from the harmonic terms into the spatial variations of the steady flow field~$\mathbf{v_2}$ is directly controlled by the viscosity of the host fluid -- in the limit of an inviscid fluid with $\eta = 0$, only the pressure gradient remains and there is no acoustic streaming.}

In order to assess the degree to which acoustic streaming plays a part in the dynamics of acoustically manipulated structures, we compute a characteristic lengthscale for acoustic streaming. Such a lengthscale can be derived from considering an infinite flat plate oscillating at frequency~$\omega$ relative to a fluid with dynamic viscosity~$\eta$ (or kinematic viscosity $\nu = \eta/\rho_m)$ and density~$\rho_m$. For the case where the oscillation direction is parallel to the plate surface, the amplitude of velocity oscillations in the fluid decays exponentially away from the plate, with a characteristic lengthscale $\delta = \sqrt{2 \nu/\omega}$.

This lengthscale, which we will refer to here as the viscous skin depth, again compares the fluid viscosity and its inertia due to the oscillation of the solid surface. Effectively, a solid surface oscillating relative to a fluid carries a viscous ``shell" with it, inside of which acoustic streaming can be strong compared to the propagation of acoustic waves from the oscillatory motion of the solid. 
Although we have considered streaming due to the presence of an oscillating wall, we emphasize that streaming arises due to any \emph{relative} oscillatory motion between a solid and a viscous fluid, such as when a solid particle is acoustically levitated~\cite{trinh1994experimental,gormley1998observation,hasegawa2009visualization}. 

The effect of acoustic streaming on the physics of an acoustically levitated object thus becomes pronounced when the characteristic size of that object is comparable to~$\delta$. 
For typical ultrasound in water using 1 MHz,~$\delta \approx 1\mu$m, or the size of a colloidal particle; for particles levitated in air  at 100kHz we have~$\delta \approx 10\mu$m. With micron-scale particles exposed to an acoustic field, forces due to acoustic streaming can thus be comparable to acoustic radiation forces, or even larger~\cite{tiwary1986hydrodynamic,spengler2003ultrasonic,Settnes2012,sepehrirahnama2016effects,pavlic2022interparticle}. For sufficiently small, dense particles, streaming can even switch the equilibrium levitation position from the pressure nodes to the antinodes~\cite{baasch2019acoustic}.
\textcolor{black}{For bubbles the interface with the surrounding medium is no longer characterized by a no-slip boundary condition. Still, streaming can have a pronounced effect on the interaction of bubbles, and may be responsible for previously mentioned repulsive bubble interactions at very close range~\cite{doinikov1995mutual,doinikov2003acoustic}.}

In the case where particles possess a high degree of symmetry, such as spheres, cylinders, or spheroids, the spatial structure of acoustic streaming around a particle can be calculated analytically~\cite{raney1954acoustical,holtsmark1954boundary,lane1955acoustical, wang1965flow,riley1966sphere,lee1990outer, zhao1999singular, rednikov2004steady}. Generally, the flow around a sphere takes the form of two sets of counter-rotating vortices: one set within the boundary layer, with characteristic size~$\delta$~\cite{hamilton2003acoustic,muller2013ultrasound} (usually referred to as inner streaming, or Schlichting streaming), and another whose spatial extent can be many times larger than the size of a particle~\cite{nyborg1958acoustic} (usually referred to as outer streaming, or Rayleigh streaming).  Controlling both these streaming forces and acoustic radiation forces has recently been shown to greatly expand the capabilities of single-particle acoustic manipulation~\cite{zhang2020acoustic,li2021three,zhu2021acoustohydrodynamic}.

\textcolor{black}{To zeroth order, the viscosity-induced effects on interactions among objects within the levitation plane can be accounted for by an effective size that simply adds an additional layer of thickness $\delta$ to the radius $a$. This is a useful approximation in the far-field limit $r \gg a, r>\lambda$ \cite{sepehrirahnama2022acoustofluidics}. 
At close approach the situation becomes more interesting. For two solid, identical spheres levitating in the pressure nodal plane, Fabre et al. showed \cite{fabre2017acoustic} that the inner vortices around the spheres generate a repulsive force that counteracts the attractive force from sound scattering (Eq.~\ref{eq:pairwise_Frad}).
The corresponding flow field around one of the spheres is shown in Fig.~\ref{fig:forces}d, where the black arrows indicate the forces experienced by the second sphere.
The degree to which viscous streaming affects the interactions observed in the inviscid case can be parameterized by the Stokes number $\Omega$ (Eq.~\ref{eq:Omega}).}
%\cite{fabre2017acoustic}}.
%
%\begin{align}
 %   \Omega = \frac{a^2 \omega}{\nu} = 2\left(\frac{a}{\delta}\right)^2.
%\end{align}
\textcolor{black}{Smaller Stokes number corresponds to increasing significance of micro-streaming. In Fig.~2 we therefore plot $1/\Omega$ along the vertical axis.}

\textcolor{black}{A direct consequence of the competing attractive scattering  and repulsive microstreaming forces is that two solid spheres no longer are driven into direct contact, but instead attain a steady-state in-plane separation that is finite.  
This happens for sufficiently small $\Omega$, below 10-20 \cite{fabre2017acoustic}, and is shown in Fig.~\ref{fig:forces}e, where the zero-crossings of the net interaction force move to larger center-to-center separations $r$ as the sphere radius $a$ decreases (darker curves).  
Experiments that levitated solid spheres ($2a < 60$ microns) in air at 30-60 kHz have observed stable pairs of spheres exhibiting finite separation distances in close agreement with the predictions by Fabre et al. as well as Lattice-Boltzmann simulations \cite{wu2022hydrodynamic}}.

Extending these calculations to derive the streaming-induced force between a general arrangement of particles is difficult, however, as such a force depends on the configurations of all particles in the flow, and is thus inherently many-body. In the case of pairs of spheres, computational and experimental results show that particles in a vibrated fluid experience a net force which causes the spheres to align such that the line connecting their centres is perpendicular to the oscillation direction. Additionally, the particles experience a mutual force which is long-range attractive but short-range repulsive~\cite{klotsa2007interaction,otto2008measurements,fabre2017acoustic}, roughly corresponding to the inner and outer streaming regions. Observations of larger particle numbers have shown chaotic spontaneous motion, driven by the collective streaming-induced flows of the cluster~\cite{voth2002ordered,pacheco2013spontaneous}. Furthermore, superposing multiple standing waves can produce complex streaming flows that generate torque on levitated particles~\cite{wang1977first,busse1981torque,lee1989near,rednikov2003behaviour,zhang2014acoustic,lamprecht2015viscous,doinikov2018acoustic,doinikov2017acoustic}. Such rotational flows introduce additional layers of complexity to the collective flows, and by extension, the many-body forces that act on levitated particles. Incorporating an understanding of these streaming flow-induced forces, and their interplay with acoustic radiation forces, remains an outstanding challenge for acoustic levitation.

\begin{figure*}
\centering
\includegraphics[width = 2.0\columnwidth]{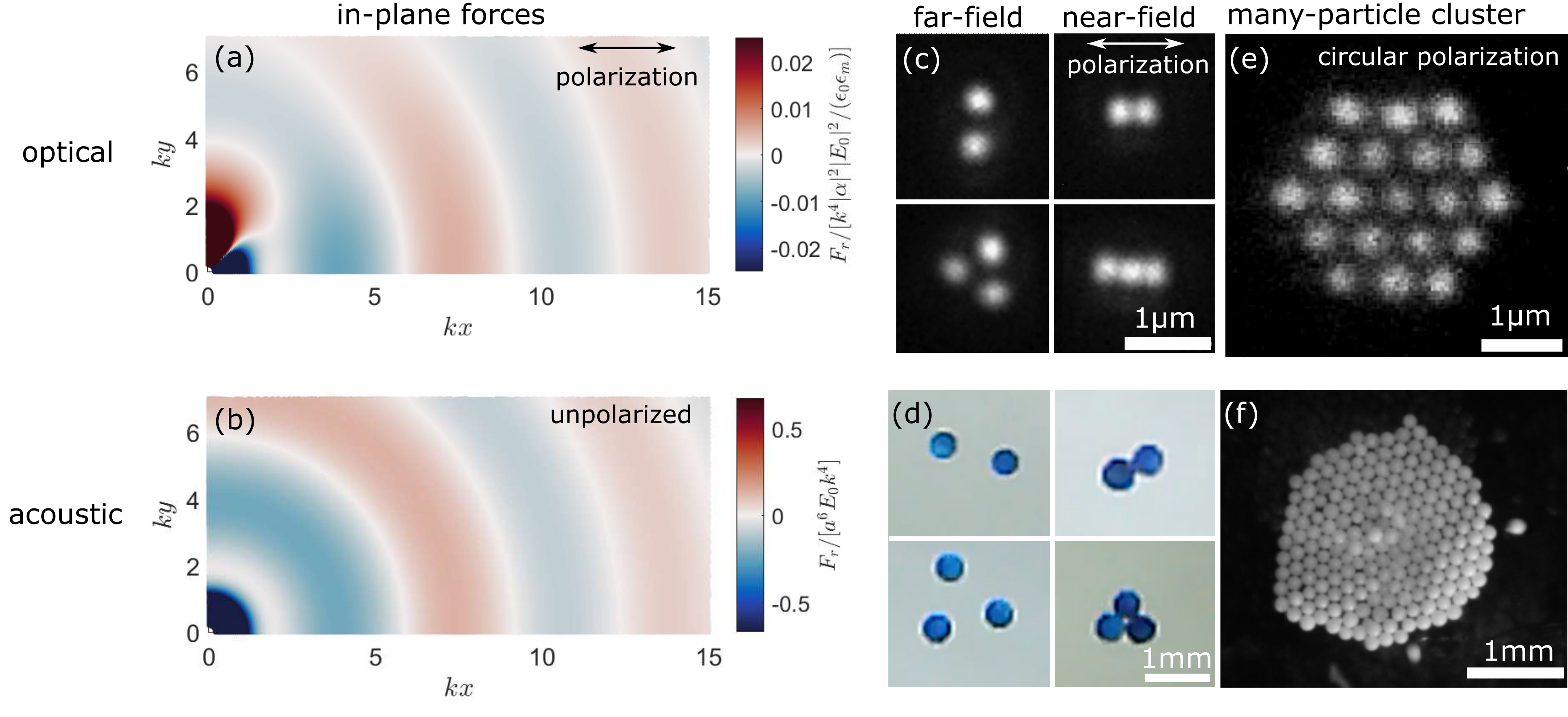}
\caption
{Comparison of optical and acoustic secondary interactions. Plots of theoretical predictions for the normalized secondary scattering radial force experienced by a particle placed in an (a) optical field~\cite{dholakia2010colloquium}, and (b) acoustic field~\cite{silva2014acoustic}, due to another particle placed at the origin (scatterers have size parameter~$ka = 0.1$, and are not visible on this scale). The incident wave is along the z-axis, and is polarized in the x-direction for the optical case. In the nodal plane, optical and acoustic secondary scattering appear similar in form near the axis of polarization. (c,d) Images of two- and three-particle clusters mediated by optical and acoustic scattering. Both the far-field binding, where particles are trapped approximately one wavelength apart, and the near-field binding are demonstrated. (c) Optical images reproduced from~\cite{yan2014potential}, showing 100nm Ag particles in water, trapped in a focused 800nm Gaussian beam (propagation direction out of the plane), with the same x-direction polarization as in (a). (d) Image of acoustic clusters from~\cite{wang2017sound}, depicting polystyrene particles in water, trapped in a 1MHz ultrasound standing wave. (e) Image of optically bound many-particle cluster, consisting of 250nm Au nanoparticles bound in far-field potential minima under a circularly polarized 800nm incident beam. Reproduced from~\cite{nan2022creating}. (f) Image of acoustically bound many-particle cluster, consisting of polyethylene particles levitated by 40kHz ultrasound in air, forming a close-packed 2D raft due to near-field attractions and, at this particle size, negligible repulsive microstreaming. Reproduced from~\cite{lim2022mechanical}. }
\label{fig:optics}
\end{figure*}

\subsection{Connections to optical radiation forces}
As Lord Rayleigh pointed out, just as the pressure of acoustic vibrations can exert forces on particles, other types of vibration must exert similar forces, under a similar mathematical framework. The most famous example is optical forces, which appear due to electromagnetic ``vibrations". Indeed, these optical forces form another (extremely widespread) approach for the noncontact manipulation of matter~\cite{ashkin1986observation,grier2003revolution}. As with acoustic radiation forces, the principle of operation relies on the scattering of a beam of light by suspended particles, which then experience an optical radiation pressure. The precise spatio-temporal force profile exerted on the suspended particles can then be tuned by shaping the beam of light. Indeed, recent advances in acoustic manipulation have recapitulated some of the library of technologies developed for advanced optical manipulation, including acoustic holography~\cite{melde2016holograms,marzo2019holographic}, controlled rotation via the transfer of orbital angular momentum~\cite{anhauser2012acoustic,jiang2016convert,marzo2018acoustic}, and acoustic tweezers~\cite{meng2019acoustic,tian2019wave,baudoin2019folding,baudoin2020spatially}.

Although the superficial similarities between acoustic and optical manipulation are very strong, there are significant differences in their underlying physics. Unlike light, whose mathematical framework is a vector field with polarization, acoustic forces fundamentally arise from the scalar (at first order) pressure, velocity, and density fields. These acoustic fields are furthermore descriptors of an underlying fluid that is dissipative and nonlinear (for example, there is no streaming equivalent for optical forces). A detailed analysis, especially for the form of the radiation pressure, can be found in several recent reviews~\cite{thomas2017acoustical,baudoin2020acoustic, dholakia2020comparing}. Here, we briefly cover some fundamentals of the optical and acoustic radiation forces. 

As with acoustic forces, the physics of optical trapping can be understood by separately considering the regimes of very small particles (size much smaller than wavelength, Rayleigh limit), and large particles (size comparable to or larger than the wavelength, Mie limit). We begin with the limit where particles are many wavelengths in size, such that light-matter interactions are well described by ray optics. Our incident wave is (coherent) light, which spreads out from a focal point, at which the area flux of photons is greatest. A dielectric particle whose diameter is much larger than the wavelength (often referred to as Mie particle) is now placed in this beam, away from the focal point. In this case, the light is refracted and reflected around the object, meaning that the momentum of the incident photons is redirected by the Mie particle. The Mie particle thus experiences an equal and opposite force towards the focal point, both parallel to and perpendicular to the beam axis~\cite{ashkin1986observation,grier2003revolution,dholakia2010colloquium,bowman2013optical}. 

Although a parallel analysis has been proposed and analyzed theoretically for ray acoustics~\cite{lee2005theoretical,zhang2011geometrical}, the range of particle sound absorbency and beam conditions that permit stable levitation appears to be very small. Additionally, current experimental techniques for acoustic levitation in the Mie limit have so far successfully levitated only objects of size smaller than three wavelengths, far from the regime where such an analysis would be appropriate~\cite{andrade2016acoustic,marzo2018acoustic,inoue2019acoustical,zehnter2021acoustic}. Nevertheless, the principle of redirected acoustic momentum flux can be used to levitate and exert forces on particles with size of order~$\lambda$. If the forward and backward acoustic momentum flux on a particle are imbalanced, either because of the particle shape~\cite{demore2014acoustic} or surface texture~\cite{stein2022shaping}, acoustically levitated particles can also experience an acoustic pulling force (``tractor beam").

In the Rayleigh limit, it is possible to make direct comparisons to the acoustic radiation forces discussed in Sec.~\ref{sec:rigid} by considering the electric dipole moment induced on the particle by the incident light~\cite{ashkin1986observation,grier2003revolution,abdelaziz2020acoustokinetics}. In order to make the comparison between the optical and acoustical Rayleigh limits clear, we briefly recapitulate the framework introduced \textcolor{black}{by Abdelaziz and Grier}~\cite{abdelaziz2020acoustokinetics}. We consider an electric field~$\textbf{E}(\textbf{r})$ whose components have amplitude and phase

\begin{align}
    E_j(\textbf{r}) = u_j(\textbf{r}) e^{i\phi_j(\textbf{r})}
    \label{eq:elec_decom}
\end{align}

For a spherical Rayleigh particle with complex electric dipole polarizability~$\alpha_e = \alpha_e' + i \alpha_e''$ and radius~$a$, this electric field exerts a force 

\begin{align}
\textbf{F}_e (\textbf{r}) &= \frac{1}{4} \alpha_e' \nabla \sum^3_{j=1} u_j^2(\textbf{r}) + \frac{1}{2} \alpha_e'' \sum^3_{j=1} u_j^2(\textbf{r}) \nabla \phi_j(\textbf{r}) \, .
\label{eq:photokinetics}
\end{align}
The first term in Eq.~\ref{eq:photokinetics} can be interpreted as a net gradient of the field intensity: just as with Mie particles, Rayleigh particles in an optical trap experience forces along intensity gradients. This force is expressed as a gradient of the electric field, and is thus strictly conservative. On the other hand, the second term is non-conservative, and is driven by phase gradients in the electric field. We note that the relative strength of the conservative and nonconservative forces is determined by the magnitudes of~$\alpha_e'$ and~$\alpha_e''$, which are functions of the scattering and absorption coefficients of the particle, and scale as~$(a/\lambda)^3$ and $(a/\lambda)^6$ respectively. For Rayleigh particles~$a \ll \lambda$, and so the conservative forces dominate. 

A similar expression for acoustics can be derived by considering a Rayleigh particle in an acoustic field, which we write using the (scalar) pressure component, decomposed similarly to Eq.~\ref{eq:elec_decom}: 

\textcolor{black}{
\begin{align}
    p(\textbf{r},t) =  u(\textbf{r}) e^{i\phi(\textbf{r})} e^{-i\omega t}
    \label{eq:press_decom}
\end{align}
}

Unlike the optical case, where the quadrupole contribution to the particle response can be neglected, particles in an acoustic field respond to applied pressure with both a dipole and quadrupole polarizability~\cite{silva2014acoustic2}. Retaining the notation of Sec.~\ref{sec:rigid}, these complex polarizabilities can be written to lowest order as 

\begin{align}
    \alpha_a &= \frac{4 \pi a^3}{3\rho_m c_m^2} f_0 \left[ -1+i\frac{1}{3}(f_0 +f_1)(ka)^3\right]\\
    &= \alpha_a'+i\alpha_a''
\end{align}
for the dipole polarizability, and 
\begin{align}
    \beta_a &= \frac{2\pi a^3}{\rho_m c_m^2}f_0\left[ 1+i \frac{1}{6} f_1 (ka)^3 \right]\\
    &= \beta_a'+i\beta_a''
\end{align}
for the quadrupole polarizability. \textcolor{black}{We note that the dipole and quadrupole polarizabilities are complex nonlinear combinations of the monopole and dipole scattering coefficients from Sec. IA. }The radiation force due to the pressure field Eq.~\ref{eq:press_decom} on such a particle can then be written as

\begin{align}
\begin{split}
    \textbf{F}_a(\textbf{r}) &= \frac{1}{4} \alpha_a' \nabla  u^2 + \frac{1}{2} \alpha_a'' u^2 \nabla \phi \\
    &+ \frac{1}{4}\beta_a'\nabla \left(u^2 +\frac{1}{2} k^{-2}\nabla^2 u^2\right)\\
    &+ \frac{1}{4}\beta_a'' k^{-2} [ (2k^2u^2+\nabla^2 u^2 + 2u\nabla u \cdot \nabla)\nabla \phi \\
    &- (u \nabla^2 \phi+2 u \nabla \phi \cdot \nabla)\nabla u ] \, .
    \end{split}
    \label{eq:acoustokinetics}
\end{align}

In the limit that the complex part of the polarizabilities is zero (i.e. the scattered acoustic field is in phase with the incident field), Eq.~\ref{eq:acoustokinetics} reduces to only the first and third terms, giving a purely conservative force and reducing to Eq.~\ref{eq:Urad} (via the definition Eq.~\ref{eq:def_vpot}). Additionally, considering only the dipole part of the particle response (the first two terms in Eq.~\ref{eq:acoustokinetics}) produces a force that looks strikingly similar to its optical counterpart, Eq.~\ref{eq:photokinetics}. Since the quadrupole polarizability depends solely on~$f_1$, which corresponds to a density mismatch between the particle and acoustic field, we conclude that the optical and acoustic forces on Rayleigh particles are identical in form when particles are completely density matched to the acoustic fluid~\cite{toftul2019acoustic}. More specifically, the quadrupolar terms in Eq.~\ref{eq:acoustokinetics} arise from the need for the velocity field to be continuous at the particle boundary, a condition that does not exist in the optical case.  

As with their optical counterparts, the conservative part of~$F_a$ scales as~$(a/\lambda)^3$, while the non-conservative parts scale as~$(a/\lambda)^6$. Thus the conservative component of the acoustic force, Eq.~\ref{eq:Urad} dominates for particles in the Rayleigh limit. However, as particles become the same size or larger than the wavelength, the nonconservative part of Eq.~\ref{eq:acoustokinetics} can become significant compared to the conservative part. Unlike the optical radiation force, the acoustic non-conservative forces (the second and fourth lines of Eq.~\ref{eq:acoustokinetics}) are not straightforwardly related to the intensity and phase gradients of the pressure field, and instead take the form of nonlinear combinations of gradients in both intensity and phase. These nonlinearities make the levitation and manipulation of Mie particles a challenging computational task. 

Just as pairs of particles placed in an acoustic trap experience interparticle interactions due to secondary scattering, pairs of particles placed in a optical trap similarly experience a force referred to as optical binding~\cite{burns1989optical,burns1990optical,tatarkova2002one, ng2005photonic,karasek2007analytical,bowman2013optical,yan2013guiding}. We consider the case of a pair of particles placed in an optical trap. One particle (the source) receives the incident field, and develops an induced dipole moment that oscillates according to its complex polarizability~$\alpha$. This oscillating dipole then produces a secondary electric field, which produces the optical binding on the second particle. Optical binding, like acoustic binding, thus produces interparticle forces that depend strongly on the orientation of the particles relative to the incident field. Unlike the acoustic version, however, optical interparticle forces also depend strongly on the orientation of the particles relative to the beam polarization. As a concrete comparison, the near-field radial force between a pair of spherical particles with radius~$a$, oriented such that their mutual axis is perpendicular to the beam direction and parallel to the beam polarization, is~\cite{dholakia2010colloquium} 

\begin{align}
    F(r) &=-\frac{3|\alpha|^2 |E_0|^2}{4\pi \epsilon_0 \epsilon_m r^4} \, ,
    \label{eq:sec_optical}
\end{align}

and in the far-field is

\begin{align}
F(r) &= \frac{|\alpha|^2 |E_0|^2 k^2}{4\pi \epsilon_0 \epsilon_m}\frac{\cos(kr)}{r^2},
    \label{eq:sec_optical_far}
\end{align}
where we have denoted the amplitude of the incident electric field as~$E_0$, which propagates in vacuum with permittivity~$\epsilon_0$, and in the trap with relative permittivity~$\epsilon_m$. This expression is once again only valid in the Rayleigh limit, where particles can be treated as point dipole sources. 

In the limit where Eqs.~\ref{eq:sec_optical} and~\ref{eq:sec_optical_far} are valid, they bear a striking similarity to the acoustic versions (Eqs.~\ref{eq:pairwise_Frad} and~\ref{eq:pairwise_Frad_far}, see Fig.~\ref{fig:optics}a,b for a graphical comparison), bearing in mind that $\alpha \sim \epsilon_0 \epsilon_m a^3$: both are strictly attractive, scale with a polarizability and scattering cross-section squared, and decay with the same power of~$r$. Here, the electric dipole polarizability plays the part of the acoustic density contrast, which contributes to the acoustic quadrupolar polarizability. 

The fact that these forms are identical stems from the fact that for Rayleigh particles with real polarizabilities in the nodal plane, the acoustic velocity field can be treated as conjugate to the optical electric field. This mathematical equivalency can be derived by treating the acoustic field using the velocity potential instead, from which the pressure and velocity fields are derived. Such an approach suggests a path for leveraging the well-developed theoretical frameworks~\cite{purcell1973scattering,draine1988discrete,marago2013optical} for optical radiation forces for their acoustic equivalents, particularly in the case of particles with nontrivial shape or acoustic resonances~\cite{kelly2003optical,noguez2007surface,amendola2017surface}. 

In practice, the shape of the incident field is often different between the optical and acoustic cases. In particular, optical traps for particles tend to be tightly focused around the particles, in order to provide good spatial localization, while acoustic traps tend to have gradients much larger than the size of a levitated particle. This distinction means that optical forces have been employed to trap smaller particles, while acoustic forces remain large for larger particles~\cite{dholakia2020comparing}. Additionally, since acoustic traps are less focused compared to typical optical traps, the shape and size of a \textcolor{black}{trapped} multi-particle cluster can differ significantly between the two methods. 

An example is illustrated in Fig.~\ref{fig:optics}c and Fig.~\ref{fig:optics}d. In the far-field, shown in the left top and bottom images in these figures, the optical and acoustic two- and three-particle clusters appear quite similar.  However, for the optical traps, the near-field binding can only be achieved by \textcolor{black}{focusing the beam more tightly}, 
%thus forcing the particles into contact. Even in this configuration,
and
the particles do not relax into the close-packed triangular cluster exhibited by the acoustic counterpart. This anisotropy is due to the laser polarization direction, which is parallel to the line formed by the three particles~\cite{yan2014potential}. Since acoustic waves have no polarization, the binding generated by acoustic scattering is intrinsically isotropic in the nodal plane. Optical forces can be made isotropic by using circularly polarized light to trap particles, allowing for the creation of more isotropic lattices (Fig.~\ref{fig:optics}e). However, creating a large lattice requires significant effort to homogenize the intensity gradients in the optical trapping plane and create phase gradient traps~\cite{nan2022creating}, whereas the wider spatial focus of acoustic traps lends itself to the production of larger clusters (Fig.~\ref{fig:optics}f).

Optical binding has been proposed as a potent tool to create arbitrary potential energy surfaces for the assembly of nanoscale particles into optical matter. Since optical beams can be readily shaped into a wide array of force profiles, both conservative and non-conservative, optical matter has been used to demonstrate tunable reaction pathways~\cite{yan2014potential}, optical epitaxial growth~\cite{huang2015optical}, particles with switchable (conservative to nonconservative) interactions~\cite{rieser2022tunable}, and non-Hermitian effects in large optical lattices~\cite{li2021non}. The development of acoustic equivalents for  these optical landscapes opens the door to realizing some of these possibilities in an acoustic system. Future work in acoustics may find the parallel between acoustics and optics to be a fruitful ground for the discovery of rich acoustic many-body physics. 

Our analysis thus far has treated the acoustic field as a purely scalar field. However, a recent body of work has shown that the acoustic field can, in some respects, also have the properties of a vector field. In particular, although the overall pressure field is a scalar quantity, acoustic waves can also produce local velocity fields -- in the coherent motion of the acoustic medium. These local fields appear as a result of evanescent waves, either due to interference or travel near a structured metasurface~\cite{shi2019observation}. When these local velocity fields rotate, the net effect is to create a vector field with intrinsic angular momentum (`spin')~\cite{toftul2019acoustic,burns2020acoustic,bliokh2019spin,bliokh2019transverse}, which can then transfer this spin degree of freedom to a probe placed in the acoustic field~\cite{shi2019observation}. Structuring this spin degree of freedom opens the door to an even wider array of possible acoustic landscapes, such as acoustic skyrmions~\cite{ge2021observation,muelas2022observation}. Combining these acoustic spin textures with acoustically induced interparticle interactions raises the intriguing possibility of creating acoustic versions of spin-matter interactions.

\label{sec:optics}

\section{Frontiers in our understanding of acoustic forces}
\subsection{Particles with arbitrary shape}

\begin{figure*}
\centering
\includegraphics[width = 2.0\columnwidth]{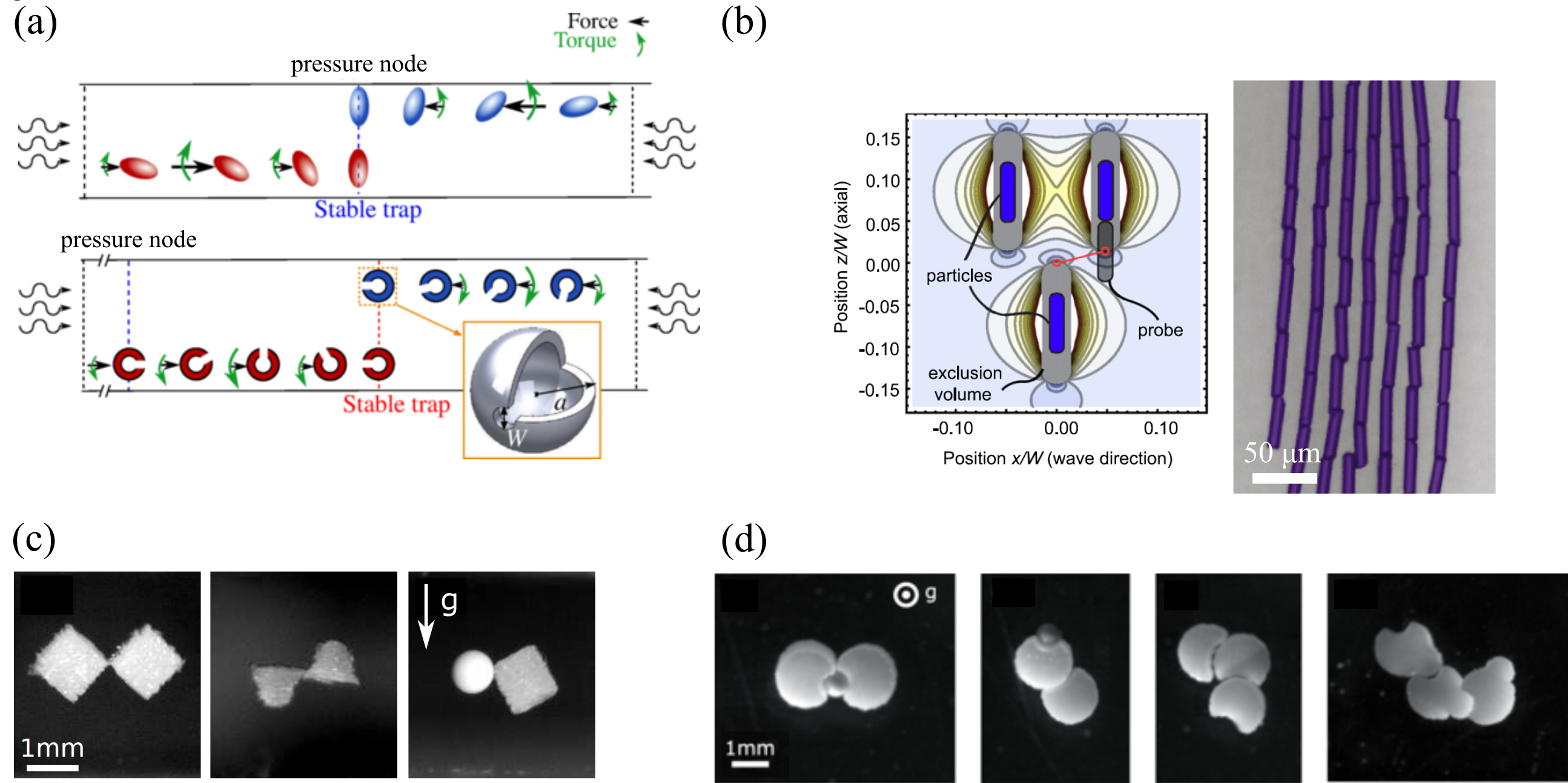}
\caption
{Particle shape alters multi-body acoustic interactions. (a) Willis coupling alters the stable levitation position and angular orientation of particles. (top) Ellipsoidal particles levitate at a pressure node and experience torques that align their long axis with the nodal plane, while (bottom) particles with nontrivial Willis coupling (shape in orange box) have stable trapping points away from the pressure node and torques that are opposite in direction (green arrows). Adapted from~\cite{sepehrirahnama2022willis}. (b) End-to-end rod alignment due to secondary scattering in a microfluidic device. (Left) Simulated acoustic potential energy for a probe rod (gray) placed in an existing particle configuration (dark blue). The nodal line is along the $z$-direction at $x$ = 0. For rods, the attractive (light blue) and repulsive (light yellow) regions are more sharply pronounced compared to spheres (see Fig. 3a, where, however, the nodal plane is at $z$ = 0). (Right) False-color optical microscope image, showing the assembly of rods into parallel columns, separated due to competition between the primary acoustic force, which drives particles toward the node along $x$ = 0, and the side-by-side repulsion of neighboring rods. Adapted from~\cite{collino2015acoustic}. 
(c) Side views of levitated particles with sharp edges. Such edges produce strongly directional bonds in  cubes, cones, and cylinders, which then act as elastic hinges when the assembled structures oscillate vertically in the acoustic field. Reproduced from~\cite{lim2019edges}.  (d) Bottom views of levitated particles with shapes designed to form lock-and-key assemblies by matching local curvature. Reproduced from~\cite{lim2023acoustically}.  }
\label{fig:shapes}
\end{figure*}

Our previous discussions of the acoustic forces (whether scattering, deformation, or streaming-dominated in Fig.~\ref{fig:regimes}) have so far focused on the case in which scatterers can be treated as spherical. This limit is useful for detailed analytical treatment, and results in the expressions for acoustic radiation force that have been previously presented. However, experimental realizations of acoustic levitation frequently utilize particles that are highly nonspherical, such as rods~\cite{wang2012autonomous}, fibres~\cite{collino2015acoustic}, muscular tissue~\cite{armstrong2018engineering}, and red blood cells~\cite{antfolk2015acoustofluidic}. In these cases, several  opportunities for new physics emerge. First, since particles are no longer isotropic and have rotational degrees of freedom, the torque exerted by the acoustic field on a single levitated particle becomes a meaningful quantity to compute. This torque determines the stable orientation of a single levitated particle. Second, since particles no longer scatter isotropically, shape modifies the form of the acoustic force acting on a particle. Finally, this modified scattering also generates modified secondary acoustic forces, resulting in shape-dependent, anisotropic acoustic interactions between levitated particles. Here, we review several recent theoretical and experimental efforts that aim to build an understanding of the effect of shape on acoustic forces and torques. 

The torque on an acoustically levitated object can be relatively straightforwardly computed by the conservation of angular momentum: the flux of angular momentum into a surface bounding the object is equal to the torque on that object (assuming that there are no sources or sinks of angular momentum inside the surface)~\cite{maidanik1958torques}.
%Quantitatively, the torque on an object placed at position vector~$\mathbf{r}$, surrounded by a bounding surface~$S$ with outward normal~$d\mathbf{S}$, subject to a sound wave with velocity field~$\mathbf{v}$ and density~$\rho_m$ is 

%\begin{align}
   % \pmb{\tau} = -\rho_m \int_S d\mathbf{S}\cdot \langle \mathbf{v(r\times v)}\rangle \, .
    %\label{eq:acrad_torque}
%\end{align}

For particular shapes with a high degree of symmetry, the expression for the torque can be simplified. 
Historically, significant effort has focused on the acoustic forces and torques exerted on disk-shaped particles, which orient themselves such that their flat faces are normal to the sound propagation direction, and enhance the primary acoustic force relative to the equivalent volume sphere~\cite{xie2004dynamics,garbin2015acoustophoresis,kepa2022acoustic}.
Additional work has highlighted the acoustic radiation forces and torques that result for other highly symmetric particle shapes, such as ellipsoids~\cite{foresti2012acoustic,silva2018acoustic,leao2020acoustic,lima2020nonlinear,lopes2020acoustic,lima2021mean}, shells~\cite{hasegawa1993acoustic,mitri2012axial,mitri2005acoustic}, and cylinders~\cite{wei2004acoustic,mitri2015acoustic}. 

In the limit of scattering particles much smaller than the wavelength, Z. Fan and coworkers~\cite{fan2008acoustic} presented a general formula for the acoustic radiation torque on a levitated particle of arbitrary shape. 
%These formulas (including the general case of Eq.~\ref{eq:acrad_torque}) take the form of surface integrals in a three-dimensional volume. 
Such an approach lends itself well to computational methods such as finite-element~\cite{muller2012numerical,glynne2013efficient,lim2023acoustically,wijaya2015numerical} or lattice-Boltzmann~\cite{barrios2008dynamics, lim2022mechanical}, but can be computationally intensive. One proposed method to decrease this computational cost is the transition matrix method (TMM), developed in analogy with the optical version of the same method~\cite{nieminen2007optical}. In this method, the incoming and scattered acoustic waves are expressed as series expansions. The scattering contribution of the object is then expressed as a transition matrix, which multiplies the coefficients of the incoming acoustic wave to give the scattered wave coefficients. Since this transition matrix is a function of the object geometry and material, this matrix can be computed or measured ahead of time, and then employed to rapidly compute the acoustic radiation force and torque. TMM has been successfully implemented and shown to be particularly efficient for objects with a high degree of symmetry~\cite{gong2019t,andrade2011matrix, gong2019reversals,hawkins2020numerical}. 

An alternative recent approach has focused on capturing the effect of shape by extending Eq.~\ref{eq:Urad} to include additional scattering coefficients. For spherical particles, the scattering coefficients~$f_0$ and~$f_1$ (in the language of Sec. ID, the acoustic dipole and quadrupole polarizabilities) couple to the pressure and velocity fields, respectively. 
Recalling that~$f_0$ corresponds to a compressibility contrast, and~$f_1$ to a  density contrast, the statement that~$f_0$ contributes to the scattered pressure and~$f_1$ to the scattered velocity is a restatement of the two constitutive relations for continuum materials: the sound momentum density is related to the velocity, and the stress in the fluid is related to the volume strain. 

However, for particles or structures that are asymmetric on a microstructural level, this assumption may be violated, such that the particle strain contributes to the sound momentum density, and the velocity to the fluid pressure. This coupling is known as Willis coupling and originated from the theoretical description of metamaterials~\cite{milton2007modifications,muhlestein2017experimental}. Within this framework, and under the assumption that the sound wave and particle obey reciprocity, the effect of a scatterer on the pressure and velocity fields can be summarized with three sets of coefficients: the scalar~$\alpha_{pp}$, which is the compressibility contribution to the pressure and for spheres is proportional to~$a^3 \rho_m \beta_m f_0$, the three-by-three tensor~$\pmb{\alpha}_{vv}$ which is the density contribution to the velocity and  for spheres is proportional to~$a^3 \rho_m f_1/\omega$ times the identity matrix, and the three-by-one vector~$\pmb{\alpha}_{pv}$, which is the Willis cross-coupling and is zero for spheres. The acoustic force on a levitated Rayleigh scatterer with these coefficients can be expressed using vectors and dyadics as~\cite{sepehrirahnama2021acoustic,sepehrirahnama2022willis} 

\begin{align}
    \mathbf{F} = &-\langle \frac{\alpha_{pp}}{2\rho_m} \nabla p^2 \rangle +  \langle i\omega \pmb{\alpha}_{vv} \mathbf{v}\cdot \nabla \mathbf{v}\rangle \nonumber \\
    &+ \langle \frac{\pmb{\alpha}_{pv}}{\rho_m}\cdot (p \nabla \mathbf{v} - \mathbf{v}\nabla p)\rangle \, ,
\end{align}
and the acoustic torque can be expressed as

\begin{align}
    \mathbf{T} = \langle i\omega p \pmb{\alpha}_{pv}\times \mathbf{v} \rangle + \langle i\omega (\pmb{\alpha}_{vv}\mathbf{v})\times \mathbf{v}\rangle \, .
\end{align}

Calculating the acoustic torque and force on an object thus reduces to the problem of computing the sets of coefficients~$\alpha_{pp}$,~$\pmb{\alpha}_{vv}$ and~$\pmb{\alpha}_{pv}$. This is generally not straightforward, as they depend highly nonlinearly on the details of the object shape~\cite{sepehrirahnama2021acoustic}. Nevertheless, this approach has been successfully used to calculate the acoustic force and torque on objects with protrusions and internal cavities~\cite{sepehrirahnama2022willis}, demonstrating that the shape of an object can control stable levitation positions, and even reverse the direction of acoustic force and torque relative to an equivalent object with no Willis coupling (Fig.~\ref{fig:shapes}a).  We note that the Willis coupling has been measured for metamaterial elements~\cite{muhlestein2017experimental,melnikov2019acoustic}, but has not yet been applied to experimental realizations of acoustically levitated particles. 

In addition to modifying the primary acoustic force and torque, particle shape can also strongly modify secondary acoustic forces and torques. Shape-dependent secondary forces have been shown to drive highly anisotropic assembly of  objects~\cite{collino2015acoustic,owens2016highly}, such as end-to-end tilings of colloidal rods (Fig. 5b). In particular, large curvature, e.g. in particles with sharp edges, appears to strongly enhance local acoustic forces, leading to the attachment of cubes along their edges instead of face-to-face ~\cite{lim2019edges} (Fig. 5c). Designing particles with appropriately matching local curvatures can also be exploited to enhance site-specific binding probabilities~\cite{lim2023acoustically} (Fig. 5d). 

Unlike the previously mentioned frameworks being developed to calculate the primary acoustic force on objects with arbitrary shape, there is currently no similar theoretical framework to treat the secondary acoustic forces. Current work on understanding the shape-dependence of the secondary acoustic force is limited to finite-element simulations, which make use of the Rayleigh-limit expressions (Eq.~\ref{eq:Urad}) to compute the force on a spherical Rayleigh scatterer due to an object of arbitrary shape~\cite{lim2019edges,lim2023acoustically}. Such a calculation scheme can provide a qualitative understanding of the structure of secondary acoustic forces, but it does not fully account for the complexity of interaction between a pair of anisotropic particles, especially since the secondary acoustic forces and torques will be a function of the position and orientation of each levitated object. Future work that seeks to elucidate these shape-dependent secondary forces thus requires the development of new experimental tools to accurately produce shaped particles, and measure their forces as a function of shape, orientation, and position, in addition to novel numerical and theoretical tools. Further opportunities for research present themselves in the Mie limit, where shape dependencies are expected to become even more extreme, or when particles with shape have some flexible elements. 

We note that our discussion of the effect of particle shape has focused only on the acoustic radiation forces and torques. However, the viscous flows around a particle are also a strong function of the particle shape, and are likely to contribute strongly to the total force and torque on an object in an acoustic field. Indeed, several studies have pointed out that the contribution of viscous torques can be very significant, even for particles much larger than the viscous skin depth~\cite{foresti2012acoustic, baresch2018orbital,pavlic2022influence}. The effect of shape on acoustic streaming-induced interaction forces remains an open frontier of research. 

\label{sec:shape}

\subsection{Energizing Instabilities}

A key assumption of our treatment of the acoustic force has been the presence of an acoustic standing wave that is weakly perturbed by the presence of a levitated object. In any experimental system for acoustic levitation or manipulation, such a standing wave is generally established by exciting a mode of an acoustic cavity (see Fig.~\ref{fig:primary} for examples). The precise shape of this mode can be controlled using the boundary conditions of the acoustic wave, such as the shape of the cavity. 

However, the levitated object forms an additional boundary condition within the acoustic cavity. Since the resultant cavity mode is also the source of forces acting on the levitated object, this fact gives rise to a class of non-conservative forces that act on levitated objects due to feedback instabilities between the levitated object and the acoustic mode present. 
Briefly, the acoustic mode present in a resonator is perturbed as the position of a levitated object changes, and this provides the possibility for positive feedback that accelerates particles.
Typically, this mechanism excites oscillations aligned with the steepest gradient of the acoustic potential (i.e.~perpendicular to the wavefronts).

The presence of instabilities in levitation systems is generically remarked upon by practitioners. Our current understanding of the effect begins with observations made in microgravity, as part of an experiment carried aboard space shuttle flight STS-41b and as reported by D. Elleman, T. Wang and M. Barmatz \cite{elleman1988acoustic}.
These authors noted that spontaneous oscillations would often occur and cause early termination of experiments due to sample ejection or contact with the boundaries of the cavity.
From this observation, a theory of feedback instability was developed \cite{rudnick1990oscillational}, which combined the influence of levitated object position on cavity modes \cite{leung1982resonance} with time-delay effects associated with the finite lifetime of cavity modes.
It was found that such  feedback instability can produce a velocity-dependent force on levitated objects, analogous to a velocity-dependent damping term.
Crucially, this damping-like term can be negative, acting to accelerate objects rather than slow them down.
The key condition for negative damping to occur is for the acoustic cavity to be excited at a frequency higher than its nearest eigenmode.
\textcolor{black}{Combining this with Eq.~\ref{eq:F_prim} and viscous drag, we arrive at a simple description for the dynamical response of an acoustically levitated object to forces driving it away from its equilibrium position at~$z=0$ \cite{andrade2014experimental}.} 
For spontaneous acceleration, the excitation frequency must be sufficiently high that negative damping overwhelms other dissipative effects (i.e.~sources of positive damping).
Experimental studies have measured and confirmed the effect and this interpretation \cite{baer2011analysis,andrade2014experimental,andrade2019experimental, hasegawa2019oscillation, lim2019cluster} (Fig.~\ref{fig:noncon}a).

The significance of this feedback instability to many-body levitation experiments is in its ability to energize systems with many degrees of freedom in a pseudo-thermal manner.
Brownian forces on acoustically manipulated particles are typically far too small to excite meaningful dynamics at room temperature, as particles are too large - 10$\mu m$ (100$\mu m$) radius particles in water (air) require on the order of a minute to diffuse one radius.
Furthermore, acoustic forces sufficiently strong to overcome surface friction (i.e.~in a surface acoustic wave geometry) or gravity confine particles at force scales far exceeding room temperature thermal agitation.
Feedback instabilities provide a mechanism for pseudo-thermal excitations that can be controlled by the applied acoustic field rather than the ambient temperature.
An example of employing  instabilities this way was presented in Ref.~\cite{lim2019cluster}, where spontaneous vertical oscillations of particles in small clusters transfer energy to in-plane degrees of freedom, exciting a variety of changes in cluster configuration (Fig.~\ref{fig:noncon}b).

Although currently lacking as cohesive of a theoretical model as for vertical oscillations, various authors have found that rotations of anisotropic objects are also spontaneously excited.
While it is possible to design the acoustic modes of a cavity to carry angular momentum \cite{zhang2022transfer}, spontaneous rotation can be observed even without such specially designed modes \cite{lim2022mechanical, jia2022size, abdelaziz2021ultrasonic}.
These excited dynamics can be exploited to measure the mechanical properties of levitated assemblies, for example by tracking the deformation and eventual break-up of a particle raft while it is spinning more and more rapidly \cite{lim2022mechanical}.

It is furthermore possible to produce an effect similar to spontaneous feedback instability, but by design.
Using ideas from the area of parametric excitation, the primary acoustic field can be modulated in time to pump the dynamics of trapped objects (i.e.~the effective spring constant which pins acoustically levitated objects in space can be harmonically varied).
This approach is even capable of different levels of excitation for particles in different acoustic minima of a standing plane wave \cite{dolev2019noncontact}.
Such a method should also be capable of pumping multi-body dynamics, analogous to the spontaneous instability.

\subsection{Nonconservative and nonpairwise forces}

\begin{figure*}
\centering
\includegraphics[width =2\columnwidth]{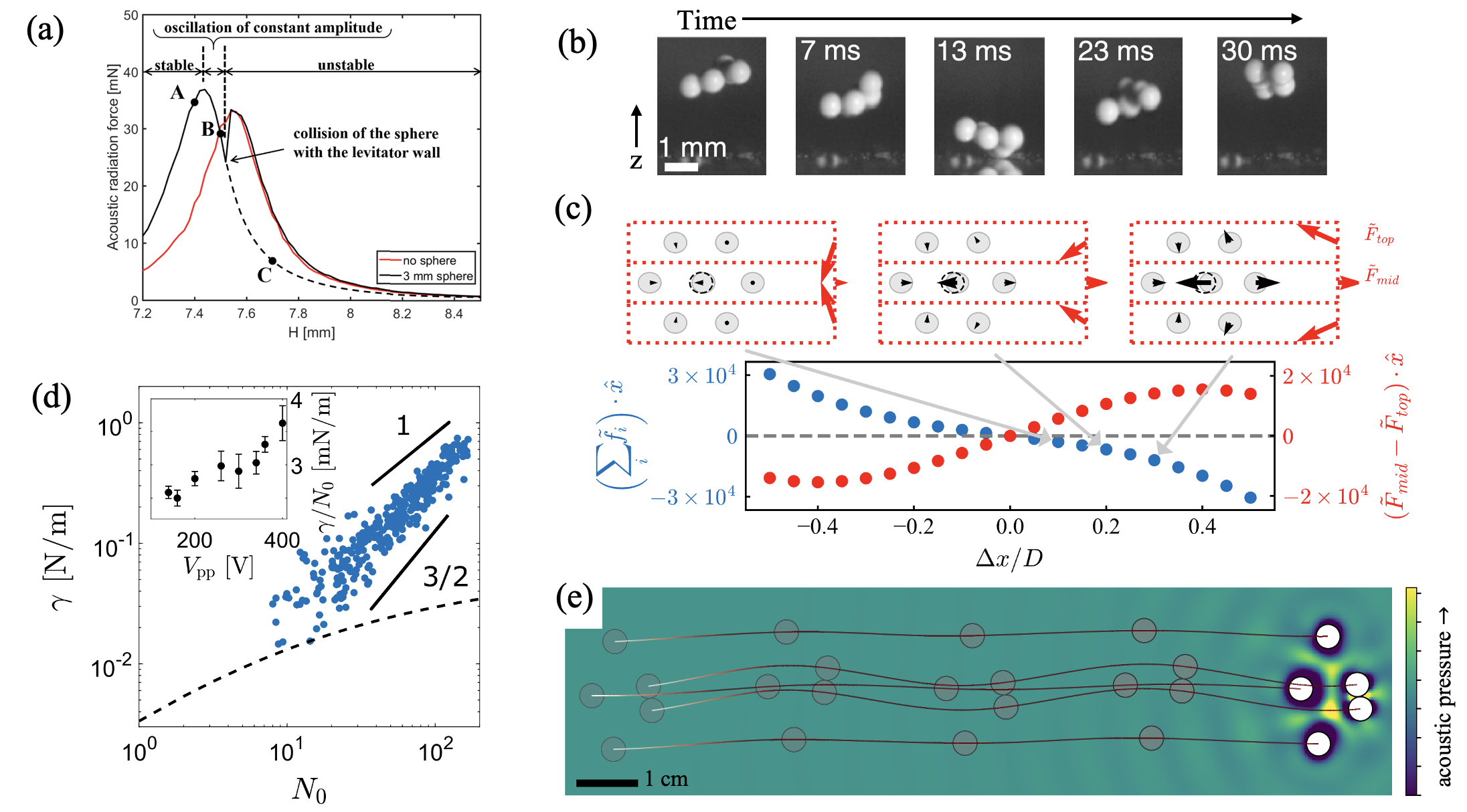}
\caption
{Non-conservative and non-pairwise effects in acoustically energized systems.
(a) Levitated objects perturb the resonant condition of a levitation cavity, as evidenced by the shift in peak force as a function of gap height H before (red curve) and after (black curve) a sphere is introduced.
Reproduced from \cite{andrade2019experimental}.
(b) The relationship between cavity mode and levitated object location can produce feedback effects that energize multi-body systems, resulting in active fluctuations that can be tuned to produce ergodic or non-ergodic reconfigurations in the levitated particle ensemble.
Reproduced from \cite{lim2019cluster}.
(c) Small particles (50$\mu$m) levitated in air experience a combination of scattering and viscous streaming interaction forces.
These forces are non-pairwise and non-reciprocal, with configuration changes (like the displacement of the center particle, shown in insets) causing net forces on the entire cluster (blue curve) as well as shearing forces (red curve).
Reproduced from \cite{wu2022hydrodynamic}.
(d) Spontaneous rotations of levitated particle rafts can be exploited to measure mechanical properties in a non-contact fashion.
Here, the effective surface tension ($\gamma$) of a levitated raft is found not only to increase with the intensity of acoustic driving (inset) but to scale with the number of particles in the raft ($N_0$). This is in contrast to the intrinsic behavior of most materials and a consequence of long-range non-pairwise interactions  (dashed line: result assuming pairwise acoustic forces).
%Effective surface tension can also be tuned by changing the intensity of acoustic driving (inset, transducer is driven by peak-to-peak voltage $V_{pp}$).
Reproduced from \cite{lim2022mechanical}.
(e) Particles of size on the order of an acoustic wavelength can display striking non-reciprocal behavior.
Small clusters of such particles experience configuration-dependent net forces, causing a spontaneous drift of the raft.
Reproduced from \cite{clair2023dynamics}.}
\label{fig:noncon}
\end{figure*}

The frameworks used to describe acoustically-driven systems (and predict their behavior) presented in Section \ref{major:one} provide valuable insights. However, they often employ an `acoustic potential' approach, resulting in conservative and pairwise descriptions of particle-particle interactions.
A growing body of literature is exploring regimes in which these properties (pairwise additivity and conservation) break down.
Note that here we are referring to an interpretation of acoustically activated systems that neglects (or coarse-grains away) the degrees of freedom of the fluid, and instead focuses on the forces present on acoustically scattering objects immersed in that fluid.
As acoustic energy is constantly injected into the fluid, it may well be more surprising that certain regimes exist in which scatterer-scatterer forces can be described by derivatives of a potential, than that they often cannot be.
Despite this, descriptions of acoustically interacting objects as interacting via forces found by the derivative of a potential have remained attractive, and extensions into more complex regimes are in their early development.
Such complications are comparatively easy to access with acoustically-interacting many-body systems, which therefore stand as promising models to advance our theoretical and practical understanding of non-pairwise, non-conservative, and non-reciprocal systems.

The non-conservative nature of acoustically-excited matter is obvious, as energy is continuously injected into the system.
The extremely helpful simplification to conservative dynamics occurs when the acoustic mode exciting the system has no net momentum, as in the case of an ideal standing wave.
Indeed, traveling waves, or waves with angular momentum, can be used to transfer momentum directly to acoustically manipulated objects \cite{abdelaziz2021dynamics,zhang2022transfer}.
In these cases, computing the primary force exerted by the acoustic mode on a point particle includes terms involving the gradient of the mode's phase (a cyclic quantity) \cite{abdelaziz2020acoustokinetics}.
This computation presumably extends to secondary scattering, i.e.~modes scattered from one particle and impinging upon another.
Extension to the two or n-body scattering problem presents substantial theoretical challenges, and so progress has mostly been demonstrated either numerically or experimentally.
Theoretical results, mainly concerning pairs of particles with differences in their material composition or geometry, have established that broken symmetry of this kind can produce unbalanced forces \cite{rajabi2018self,sepehrirahnama2022acoustofluidics}.
Air bubbles in particular can be used to create propulsion this way, for example by breaking their symmetry by encapsulating most of their surface \cite{dijkink2006acoustic, mcneill2020wafer}.
The large volume changes bubbles experience when driven near their resonance frequency enable intense microstreaming flows.
Even without such engineered complexities, bubbles can display distinct non-pairwise additive interactions due to their intense nonlinear scattering.
With as few as three bubbles, simulation studies have shown that the pair force description is insufficient \cite{chen2020modulation}.
Even the forces between a single pair of bubbles will be modulated by the presence of a third nearby bubble, particularly when the modulating bubble is super-resonant under the excitation frequency.

Intense microstreaming flows are not unique to bubbles. They can also arise in the vicinity of solid objects scattering intense sound.
Forces arising from microstreaming compete with scattering interactions between objects, particularly when the Stokes number, $\Omega=\omega a^2/\nu$ is small or, equivalently, when the particle radius is similar in scale to the viscous boundary skin depth, $\delta = \sqrt{2\nu/\omega}$.
Acoustic forces between particles in the same plane, which are attractive at short ranges for particles $a\gg \delta$, can instead display a stable fixed point at finite separation \textcolor{black}{(Fig. 3e)}, leading to `expanded' assemblies of particles which do not come into surface to surface contact \cite{wu2022hydrodynamic, thomas2004structures, fabre2017acoustic}.
The lack of (frictional) surface contact, combined with significant microstreaming contributions to interparticle forces, allows for new, non-conservative effects to become apparent.
Small ($< 50 \mu m$) particles levitated in air form hexagonal lattices with significant particle separations when driven with ultrasound.
However this system also displays spontaneous excitations, which take the form of string-like rearrangements of particles with avalanche-like bursts of motion \cite{wu2022hydrodynamic}.
Note that these excitations are not accompanied by vertical oscillations, as in systems energized by the feedback instability discussed in the previous section.
Instead, this mechanism of energy injection depends upon the presence of microstreaming flows near particles.
Numerical evidence clearly shows that such flows produce forces on particles which are neither pairwise additive, nor reciprocal (i.e.~the force from particle A on particle B is not opposite and equal to the force from B on A) \cite{wu2022hydrodynamic} (Fig.~\ref{fig:noncon}c).

While microstreaming flows can introduce intriguing complexity to interactions between acoustically-manipulated objects, under some conditions scattering forces can also elicit non-pairwise or non-reciprocal behaviors.
Groups of hundreds of particles much larger than the viscous boundary layer depth form close-packed monolayer rafts when levitated in air.
By observing spontaneous angular accelerations, the mechanical properties of such rafts can be probed in a non-contact manner \cite{lim2022mechanical}.
Surprisingly,  properties such as \textcolor{black}{an effective} raft surface tension are found to scale with the size of the raft (Fig.~\ref{fig:noncon}d), in stark contrast to the behavior of \textcolor{black}{molecular} liquids for which material properties quickly saturate with a small number of \textcolor{black}{constituents}.
This `extrinsic' scaling is indication of non-pairwise additive effects which scale particle-particle acoustic interactions as the number of nearby neighbors grows.
If particle size is increased even further, to $a\approx \lambda$, striking geometry-dependent non-reciprocal effects can  be observed.
Clusters of a few particles can adopt asymmetric configurations with net unbalanced forces, which cause the entire cluster to translate as well as excite internal vibrational modes (Fig.~\ref{fig:noncon}e) \cite{clair2023dynamics}.
Particles in such configurations do not form action-reaction pairs, and so present a rich space in which to explore the consequences and origins of non-reciprocal behavior in fluid-immersed many-body systems.

\section{Applications}

\begin{figure*}
\centering
\includegraphics[width =1.5\columnwidth]{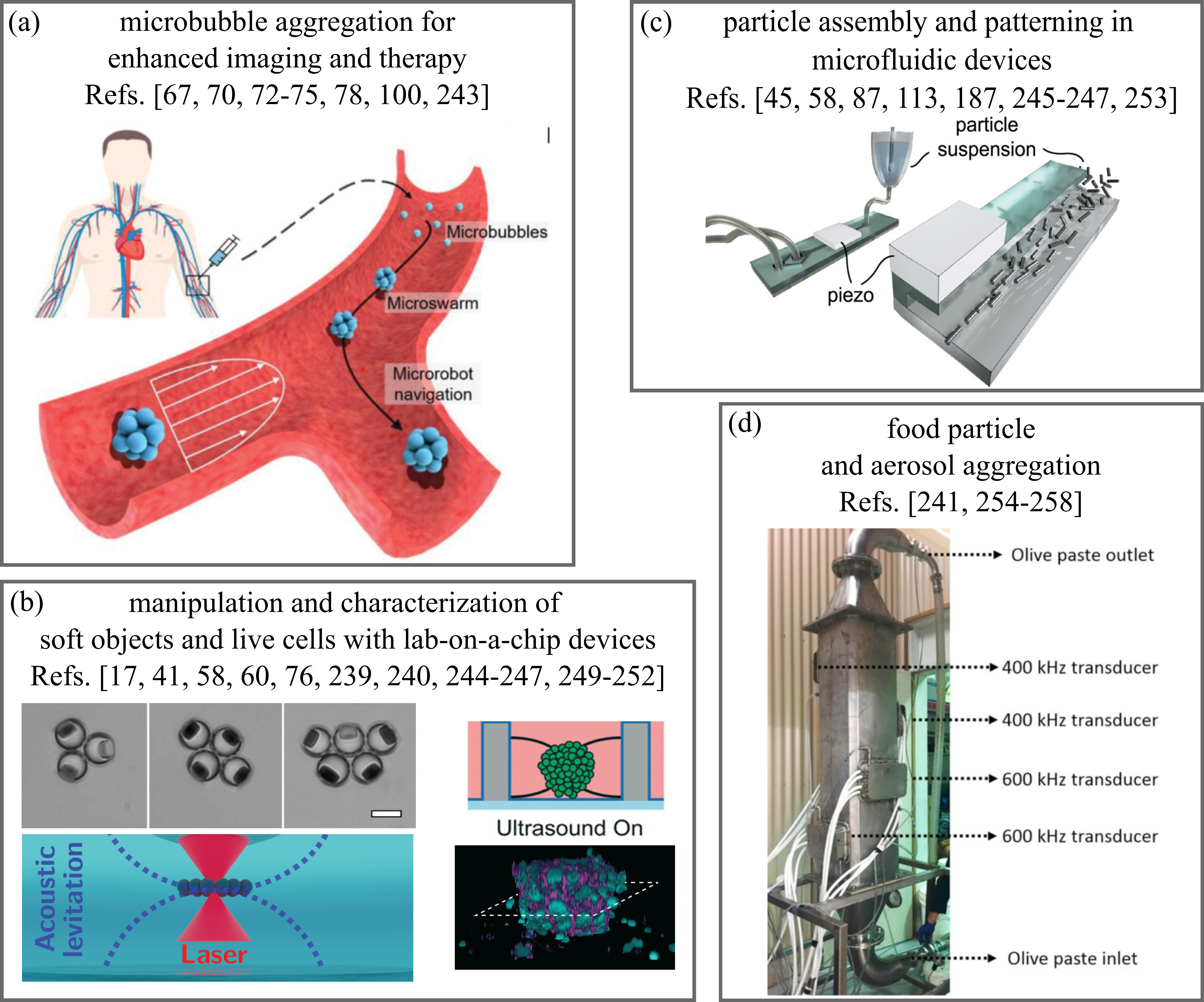}
\caption
{Applications utilizing sound-mediated interparticle forces.
(a) Microbubble aggregation for enhanced in-vivo imaging and therapy. Image:  aggregation of microbubbles into a `swarm' that functions as a sound-controlled microrobot for cargo delivery. Adapted from Ref. \cite{fonseca2022ultrasound}. 
(b) Manipulation and characterization of soft biomedical materials with lab-on-a-chip devices. Images: (top left) aggregation of emulsion droplets with orientation control of internal anisotropic cargo (discs) (adapted from Ref. \cite{shakya2022acoustically}, scale bar: 10 $\mu$m); (botttom left) in-situ Raman measurement of aggregated cells or particles (adapted from Ref. \cite{santos2021printed}); (right) sketch and confocal image of acoustically aggregated multicellular tumor spheroids (adapted from Ref. ~\cite{olofsson2018acoustic}).
(c) Particle assembly in microfluidic devices.  Image: device for aligning and patterning anisotropic particles. Adapted from Ref. \cite{collino2015acoustic}.
(d) Food particle and aerosol aggregation. Image:  megasonic reactor for olive paste aggregation. Adapted from Ref.~\cite{juliano2017extraction}.
%Applications utilizing sound-mediated interparticle forces. Boxes clockwise from top left: (A) Manipulation and characterization of soft biomedical materials with lab-on-a-chip devices. Images: (top left) aggregation of emulsion droplets with orientation control of internal anisotropic cargo (discs) [adapted from Ref. \cite{shakya2022acoustically}, scale bar: 10 $\mu$m]; in-situ Raman measurement of aggregated cells or particles [from Ref. \cite{santos2021printed}]; \textcolor{black}{cluster of XXXXXXXXXXXX; adapted from Ref.[YYY]}. (B) Particle assembly in microfluidic devices.  Image shows device for aligning and patterning anisotropic particles [adapted from Ref. \cite{collino2015acoustic}]. (C) Food particle and aerosol aggregation. Image shows megasonic reactor for olive paste aggregation [from Ref. ~\cite{juliano2017extraction}].(D) Microbubble aggregation for enhanced in-vivo imaging and therapy. Image shows aggregation of microbubbles into a `swarm' that functions as a sound-controlled microrobot for cargo delivery [adaptedfrom Ref. \cite{fonseca2022ultrasound}]. In the center of this figure the state diagram of Fig. 2 is reproduced with dotted ovals representing typical regimes covered by applications (A)-(D).
}
\label{fig:app}
\end{figure*}

As we outlined in the preceding sections, secondary acoustic radiation forces together with acoustic streaming control the interaction between objects that are levitated or moved by primary acoustic forces.  
This interaction has been exploited across a wide range of applications in which small objects are aggregated into larger clusters or assembled into patterns, either inside a liquid medium or in air.
\textcolor{black}{Using standing plane waves, these applications rely on the primary acoustic force to drive particles, for given acoustic contrast $\Phi$ (Eq.~\ref{eq:Phi}), into a pressure node or anti-node according to Eq.~\ref{eq:F_prim} and similarly drive bubbles to nodes or anti-nodes depending on whether their resonance frequency, which scales inversely with bubble size (Eq.\ref{eq:bubble-resonance}), is higher or lower than the applied sound frequency (Eq.~\ref{eq:bjerk_primary}; but see \cite{doinikov2003acoustic} for the limit of large dissipation). In mixtures of particles with positive and negative acoustic contrast this allows for separation into  nodal/anti-nodal planes or lines \cite{shakya2022acoustically,doinikov2003acoustic,silva2014acoustic, kothapalli2016investigation}.
Within such nodes, furthermore, particles attract with the secondary force that depends on their compressibility ratio or difference in material density (Eqs.~\ref{eq:pairwise_Frad},~\ref{eq:pairwise_Frad_antinode}), while bubbles can also repel within a certain frequency window (Eq.~\ref{eq:bjerk_secondary})}.
A particular advantage of using sound is that the acoustic forces can be sufficiently large to manipulate objects of almost any material type, density or shape in size from the nanoscale to 100s of microns or larger, and to configure them into arrangements that can reach the centimeter scale.  
\textcolor{black}{In Fig.~\ref{fig:app} we highlight a few groups of applications that closely relate to the regimes of many-particle interactions introduced earlier}.

The aggregation of micro-bubbles, each a few microns in diameter, via sound-induced Bjerknes forces has important benefits for medical and therapeutic applications (Fig.~\ref{fig:app}a). Bubble clusters are an effective means to enhance imaging contrast \cite{dayton1997preliminary, lazarus2017clustering}.  
Importantly, mobile acoustically bound bubble clusters were observed to remain intact at physiological flow rates \cite{dayton1997preliminary, fonseca2022ultrasound}. 
Bjerknes forces also attract bubbles to surfaces, and this has been proposed for targeted adhesion to specific sites in vessels or tissues, particularly for drug delivery applications \cite{garbin2011unbinding, kokhuis2013secondary, navarro2022monodispersity}. 
Cargo transport by ultrasonically excited bubble trains in confinement has been demonstrated \cite{janiak2023acoustic}.
More recently, the development of real-time programmable force fields using multiple ultrasonic transducers has made it possible to steer micro-bubble clusters, demonstrating the successful combination of directed self-assembly via secondary acoustic forces with controlled navigation using primary acoustic forces \cite{fonseca2022ultrasound, schrage2023ultrasound}. 
This opens up the possibility for using such clusters as acoustically controllable `microrobots’ inside a living organism.

Embedding surface acoustic wave (SAW) generators into microfabricated structures has enabled microfluidic or whole lab-on-a-chip devices that can arrange small objects into patterns that are spatially and temporally controllable (Fig. \ref{fig:app}b,c. 
Such devices, which can be 3D-printed \cite{santos2021printed}, are integrated straightforwardly into setups employing various types of microscopy. 
In their simplest form the devices send sound from opposing sides across a fluidic channel or reservoir to form a 1D standing pressure wave, which drives objects inside the channel to aggregate in the pressure nodes or anti-nodes, depending on the acoustic contrast \cite{habibi2017trapping} (see Eqs.~\ref{eq:pairwise_Frad} and~\ref{eq:pairwise_Frad_far} for rigid particles and Eq.~\ref{eq:bjerk_secondary} for bubbles in a liquid). 
With additional sets of SAW generators more complicated nodal patterns can be generated by superimposing multiple 1D standing waves along different angles \cite{gao2022staged, silva2019particle, tahmasebipour2020toward, yang2022harmonic}(Fig.1c). 
Objects can be trapped (or released) on command by turning on (or off) piezoelectric transducers that generate the primary acoustic force field, while secondary acoustic forces due to sound scattering bring the objects into close proximity \cite{chen2016onset,gao2022acoustic,kothapalli2016investigation}.  

One important application has been to manipulate and characterize soft objects such as live cells and even small organisms (Fig.\ref{fig:app}b). For example, these systems have been used for concentrating cells to enhance harvesting efficiency during sedimentation \cite{coakley2000analytical}, to probe cell-cell interactions \cite{chen2015two, saeidi2020quantitative}, or for separation or trapping of cells \cite{gao2022acoustic, kothapalli2016investigation, silva2019particle}.  
Concentrating cells within the levitation plane then facilitates investigation with microscopy and associated techniques such as Raman spectroscopy \cite{santos2021printed}, or to mimic the arrangement of cells within a tumor~\cite{olofsson2018acoustic}.
Secondary acoustic forces have also been used to arrange liquid emulsion droplets into close-packed clusters and, at the same time, manipulate the relative orientations of small anisotropic cargo inside the droplets, such as small rigid discs \cite{shakya2022acoustically}. 
While the Bjerknes forces between microbubbles tend to be in the nano-Newton range and thus quite small \cite{garcia2014experimental}, they are nevertheless sufficiently strong to bind organisms such as \emph{C. elegans} worms to individual oscillating microbubbles that are tethered to a surface \cite{xu2013microbubble}.  
Such tethering can be achieved when the bubbles sit in small wells that have been etched into the surface of a microfluidic channel. 
Large arrays of tethered microbubbles have been used not only to amplify the attractions but also to align objects whose surfaces have a suitably matching bubble pattern. When two such bubble-decorated surfaces come within the range of the Bjerknes forces, they will align laterally, which makes it possible to position flat centimeter-scale surfaces with an accuracy of a few microns simply by turning on a sound field \cite{goyal2022amplification}. 

Microfluidic devices have also been used to organize assemblies of particles into specific patterns  (Fig.\ref{fig:app}c). With a single pair of ultrasound transducers on the sides of a microfluidic channel, particles will be attracted by secondary acoustic forces to form chains or columns. 
This has been used to assemble spheres into `chains of pearls' and form columns from highly anisotropic particles such as micro-rods, which link up end-to-end (Fig. 5b) \cite{baasch2022gap, collino2015acoustic}.  
Due to the competition between the primary acoustic force, which attracts neighboring chains of rigid particles toward the nodal line, and the secondary force, which is repulsive in the direction perpendicular to that nodal line, dense patterns of closely spaced parallel chains can be formed in the vicinity of the node (see Fig. 5b), which is of potential interest for filtration applications \cite{collino2015acoustic}. 

With more complicated transducer arrangements, such as in Fig.~1c, superposition of standing sound waves leads to patterns of wells in which particles can be first trapped and then aggregated by secondary forces \cite{gao2022staged, silva2019particle, tahmasebipour2020toward, yang2022harmonic}.  In addition, with computer-controlled changes of the amplitudes and frequencies of the superimposed waves, the local particle configuration and orientation can be controlled \cite{yang2022harmonic}.

In 3D acoustic cavities, where the standing wave nodes generated by plane waves form 2D planes, attractive secondary forces have been used to generate crystal-like, close-packed clusters and monolayer particle rafts in either liquid or air \cite{garcia2014experimental,owens2016highly,santos2021printed,lim2023acoustically, lim2019cluster, lim2019edges,lim2022mechanical}. These forces also direct the assembly of particles of non-spherical shape by exploiting the local particle curvature (see Section II.A and Fig.~\ref{fig:shapes}).

While levitating monolayer rafts are confined to a nodal plane by the primary acoustic force, they are effectively unconstrained to rotate in that plane, hindered only by drag from the surrounding fluid.
For rafts levitated in air the minimal effect from drag opens up possibilities for use as a contactless rheometer, where the raft defects and shape changes can be tracked at the individual particle level while the rotation speed is increased, allowing for detailed  examination of material failure \cite{lim2022mechanical}.  

From the oscillating term in Eq.~\ref{eq:pairwise_Frad_far} we see that the secondary acoustic force exhibits additional in-plane positions where the particles are acoustically bound across distances of approximately one sound wavelength from each other, albeit more weakly than at close approach. This has been used to generate 2D lattices of well-separated spherical objects within the nodal plane \cite{wang2017sound, rabaud2011acoustically, zhang2016acoustically}. 

On much larger scales, \textcolor{black}{differences in acoustic contrast $\Phi$ together with} ultrasound-generated attractions between particles have been applied for the separation and aggregation of food particles in liquids or solid/liquid mixtures \cite{leong2015megasonic, leong2014design} as well as for stratification of sludge particles during ultrasound-assisted hot air convective drying \cite{mou2021experimental} (Fig.\ref{fig:app}d).  
Another such `megasonic’ application has been for the large scale accumulation and aggregation of aerosol particles to aid their collection and removal \cite{qiao2017aerosol, qiao2021particulate}.

Finally, there may also be applications where coupling to  acoustic forces needs to be suppressed.
\textcolor{black}{From Eq.~\ref{eq:F_prim} we see that vanishing acoustic contrast $\Phi = 0$ leaves particles unaffected by the primary force, i.e., it eliminates acoustophoresis. 
Interestingly, for plane standing waves, such acoustic `transparency' with respect to the primary force does not necessarily imply that secondary scattering interactions also vanish. This is because these interactions depend only on $f_0$ or $f_1$ (see  Eqs.~\ref{eq:pairwise_Frad} and \ref{eq:pairwise_Frad_antinode}), while $\Phi$ involves both $f_0$ and $f_1$ in combination.
Thus, particles with $\Phi = 0$  can still scatter sound and interact with other objects, for example levitated objects that exhibit a nonzero $\Phi$. }
However,  by considering special core-shell particles with an outer `cloaking' layer designed to suppress scattered sound, simulations have shown that it should be possible to effectively also eliminate the secondary radiation forces between two or more such particles and thus render them acoustically `invisible' \cite{leao2016core}.

\section{Conclusions and Outlook}
In this article we discussed acoustic levitation as a research platform to manipulate systems of multiple interacting particles. We specifically focused on the regime where the particles are spaced closely, a couple particle diameters or less apart, and where the particles themselves are much smaller than the sound wavelength (Rayleigh limit).  In this regime secondary scattering forces in concert with forces due to viscous micro-streaming give rise to a host of opportunities to tailor strong particle-particle interactions \emph{in situ}.   

Given the ability of sound to interact with objects of any shape and essentially any material, this regime has become of interest for directed particle assembly and has led to the emergence of applications ranging from microfluidic devices for controlled particle aggregation to `megasonic’ equipment for large-scale aerosol removal.
At the same time, the tunability of the sound-matter interactions makes this regime an exquisite laboratory for exploring many-body physics at room temperature under ambient conditions.
Both conservative as well as non-conservative forces can be introduced systematically, whereby the former control the steady-state particle configurations and the latter the strength of fluctuations around those configurations.
Here acoustic levitation in air offers an advantage over related systems such as colloids or dusty plasmas in that it becomes possible to access underdamped many-particle motions without Coulomb interactions. Acoustic levitation enables active fluctuations that can arise either from the feedback between particle movement and the resonant modes in the acoustic cavity or directly from the hydrodynamic coupling between moving particles.
With increasing strength they can drive levitating particle assemblies reversibly between quiescent, highly ordered steady-states and agitated disordered configurations behaving like liquids.
%Another aspect related to underdamped particle movement in the low-viscosity environment of air is that eigenmodes of particle assemblies are have well-defined resonance peaks, which allows for parametric excitation. 
Furthermore, the low-viscosity environment of air can permit particle assemblies to display long-lived vibrational modes that may be amenable to parametric excitation. 

Much of this rich dynamical behavior has only started to be explored. While mean-field models are available for dilute assemblies of point-like scatterers, new theoretical approaches are needed to properly describe the physics that emerges for dense configurations of finite-size particles, where approximations based on single scattering events may no longer hold and recent experiments as well as simulations indicate that pairwise additivity of particle-particle forces breaks down.
Some of the signatures of the underlying non-reciprocal interactions have been reported, although a fuller understanding is needed of the conditions under which they arise in acoustic levitation.
Finally, there are new and so far under-explored opportunities for tailoring particle interactions by designing specific non-spherical shapes or non-uniform material properties of the particles. This includes Willis coupling for particles or structures that are not symmetric or particles with highly flexible components.
In general, shape effects due to scattering can be expected to become even more pronounced in the Mie limit, while the effect of particle shape on acoustic streaming-induced interaction forces remains an open frontier.

On the experimental side, acoustic levitation provides an environment in which many-body physics is accessible with particles in a size range, from a few microns on up, that enables straightforward observation and tracking of individual objects inside larger assemblies with standard light microscopy (in the case of microfluidic systems) or high-speed video imaging (for levitation in air).  Especially for levitation in air, the open sides and the large size of the acoustic cavity containing the sound field provide easy access to freely floating structures in the nodal plane(s). This also allows for the straightforward application of additional electric or magnet fields or for controlled changes in the (chemical) environment, and it also opens up interesting possibilities for \emph{in situ} mechanical access via computer-controlled micro-manipulators. Finally, the large particle sizes and extremely wide range of different materials compatible with acoustic levitation open up unique new possibilities for tailoring the interactions among objects often difficult to control by other means. This includes interactions between particles made from high-density material that would sediment too quickly in a liquid but can be  levitated by the primary acoustic force, interactions among particles with designed shapes of arbitrary complexity that can nowadays be 3D-printed given that the resolution of   additive manufacturing methods has become sufficiently high, as well as interactions among biological and thus inherently active objects such as small live organisms.

\section{Acknowledgements}

We thank Tali Khain, Brady Wu, Qinghao Mao, Jason Kim, and Thomas Witten for inspiring discussions. This work was supported by the National Science Foundation through award number DMR-2104733. B.V.S.~acknowledges support through a Kadanoff-Rice fellowship from the Chicago MRSEC, which is funded by the National
Science Foundation (DMR-2011854).

%\bibliography{thebib}

\end{document}